\documentclass[superscriptaddress,nofootinbib, amsmath,amssymb,preprintnumbers,prdfloatfix,twocolumn,PRL]{revtex4-2}
\usepackage{CJK}
\usepackage{graphicx}% Include figure files
\usepackage{dcolumn}% Align table columns on decimal point
\usepackage{upgreek}
\usepackage{amsmath}
\usepackage{bm}% bold math
\usepackage[colorlinks=true,pdfstartview=FitV,breaklinks=true]{hyperref}
\usepackage[dvipsnames,table]{xcolor}
\hypersetup{urlcolor=BlueViolet,
	    citecolor=Plum,
	    linkcolor=PineGreen}
\usepackage{lipsum}	    
\usepackage{titlesec}
\usepackage{etoolbox}% http://ctan.org/pkg/etoolbox
\usepackage[normalem]{ulem}
\usepackage{tabularx}
\usepackage{amssymb}
\usepackage{float}	
\usepackage{caption}
\usepackage{subcaption}
\usepackage{booktabs}
\usepackage{comment}
\usepackage{appendix}
\usepackage{aasmacros}
\usepackage{subcaption}
 \usepackage{lettrine,Typocaps,Kinigcap,AnnSton,Kramer,Starburst,Elzevier}

\usepackage{relsize}

\newcommand{\fdm}{f_{\rm DM}}

\newcommand{\fpdm}{f_{\psi,\rm dm}}
\newcommand{\dd}{\mathrm{d}}

\begin{document}

\title{Macroscopic dark matter constraints for extended mass functions}

\author{Zachary S. C. Picker}
\email{zachary.picker@queensu.ca}

\affiliation{Department of Physics, Engineering and Astronomy, Queen's University, Kingston ON K7L 3N6, Canada}
\affiliation{Arthur B. McDonald Canadian Astroparticle Physics Research Institute, Kingston ON K7L 3N6, Canada}
\affiliation{Perimeter Institute for Theoretical Physics, Waterloo, ON, N2L2Y5, Canada}
\author{Andrew Buchanan}
\affiliation{Department of Physics, Engineering and Astronomy, Queen's University, Kingston ON K7L 3N6, Canada}
\affiliation{Arthur B. McDonald Canadian Astroparticle Physics Research Institute, Kingston ON K7L 3N6, Canada}
\author{Melissa Diamond}
\affiliation{Arthur B. McDonald Canadian Astroparticle Physics Research Institute, Kingston ON K7L 3N6, Canada}
\affiliation{Department of Physics, McGill University, Montreal, QC, H3A 0G4, Canada}
\author{Joseph Bramante}
\affiliation{Department of Physics, Engineering and Astronomy, Queen's University, Kingston ON K7L 3N6, Canada}
\affiliation{Arthur B. McDonald Canadian Astroparticle Physics Research Institute, Kingston ON K7L 3N6, Canada}

\begin{abstract}
\noindent We compile and recompute a subset of the landscape of constraints for macroscopic dark matter (macros), explicitly including the dependence on the fraction of dark matter in macros $\fdm$. This allows us to map these constraints to scenarios with extended mass functions for the macros, which is a generic outcome for many of the assembly processes relevant to macro formation. We compute a number of examples of constraints for extended mass functions, finding that in many cases, wider mass functions can be severely constrained by just the interactions of the tails of the distributions. Finally, we present the code and data required for computing these limits and generating these plots in a publicly accessible repository.
\end{abstract}

\maketitle

\section{Introduction}
\noindent \lettrine[lines=4,findent=3pt]{M}{acroscopic} dark matter can go by many names; macros~\cite{jacobs_macro_2015}, ultraheavy~\cite{kolb_wimpzillas_1998,chung_nonthermal_1998,kuzmin_ultra-high_1998}, composite~\cite{bagnasco_detecting_1994}, Compact UltraDense Objects (CUDOs)~\cite{,rafelski_compact_2012,rafelski_compact_2013}, MAssive Compact Halo Objects (MACHOs)~\cite{griest_galactic_1991}, nuggets of various kinds~\cite{witten_cosmic_1984,de_rujula_nuclearitesnovel_1984,farhi_strange_1984,alcock_evaporation_1985}, Q-balls~\cite{coleman_q-balls_1985,kusenko_supersymmetric_1998}, Fermi balls~\cite{lee_fermion_1987,lynn_strange_1990,macpherson_biased_1995}, Strongly Interacting Massive Particles (SIMPs) \cite{Starkman:1990nj}, etcetera, depending on their formation mechanism and the predilections of individual researchers. Interest in this regime has continued to grow in recent years, with a large number of macroscopic dark matter models emerging in the literature~\cite{zhitnitsky_nonbaryonic_2003,zhitnitsky_cold_2006,foadi_technicolor_2009,kribs_quirky_2010,detmold_dark_2014,krnjaic_big_2015,wise_stable_2014,wise_yukawa_2015,hardy_big_2015,gresham_nuclear_2017,gresham_early_2018,ge_cosmological_2018,bai_dark_2019,bramante_saturated_2018,SinghSidhu:2019nmh,Bramante:2021dyx,flores_primordial_2021,acevedo_loosely_2024,lu_black_2025,bramante_very_2026,Espriu:2026jzi}. In this paper we compile and recompute a wide range of the constraints in the literature for macroscopic dark matter, recasting them for the first time to include their dependence on $\fdm$, the fraction of the total dark matter energy density in macros. 
% Such a framing is familiar within the primordial black hole community, and indeed by fixing a mass-cross-section relation for the dark matter, we can produce similar plots of the $\MDm-\fdm$ parameter space by limiting ourselves to contours of the given relation. 

From a model-agnostic perspective, we can define the macroscopic regime with a simple test: if the dark matter is sufficiently massive that one would not expect it to pass through meter-scale direct detection experiments, then we consider it macroscopic. As it turns out, this regime begins around the Planck mass and extends to arbitrarily high masses. However, parameter space above $10^{-10}~M_\odot$ is well-studied by microlensing surveys, particularly in the context of primordial black holes (PBHs)~\cite{pbh,Hawking:1971ei,Carr:1974nx,Chapline:1975ojl}. In the interest of covering new ground, we therefore restrict our study here to the landscape of dark matter with masses between 1 GeV and $10^{50}$ GeV $\simeq 10^{-7} M_\odot$. We include such comparatively low masses for a simple reason---if macros are composite objects, then there may be a remaining population of the particles which comprise the composite which are either unbound, or in smaller than average composites. When computing constraints, then, we must be able to account for this population, which may yet show up as heavy individual particles in traditional direct detection experiments~\cite{akerib_search_nodate}.

Moreover, one purpose of this work is to allow for the computation of constraints in scenarios where the dark matter is described by extended mass distributions in general. This is indeed a common outcome when composite dark matter states are constructed by some process in the early universe. A procedure for transforming `monochromatic' constraints to extended mass functions is already well-known for PBHs~\cite{carr_primordial_2017,bellomo_primordial_2018}, which we adopt and extend here.

In Sec.~\ref{sec:constraints} we very briefly summarize the landscape of constraints that we include in our limits, including a detailed re-analysis of the white dwarf supernovae constraints. Then in Sec.~\ref{sec:EMF} we show how these limits can be recast as limits on extended mass functions, computing explicitly a number of plausible examples. In App.~\ref{sec:app}, we present in detail the code and data used to both compute the constraints and generate the figures in this paper. This code can be freely accessed at Ref.~\cite{macrolimits}.

\section{Constraints}\label{sec:constraints}

\begin{figure*}
    \centering
    \includegraphics[width=\linewidth]{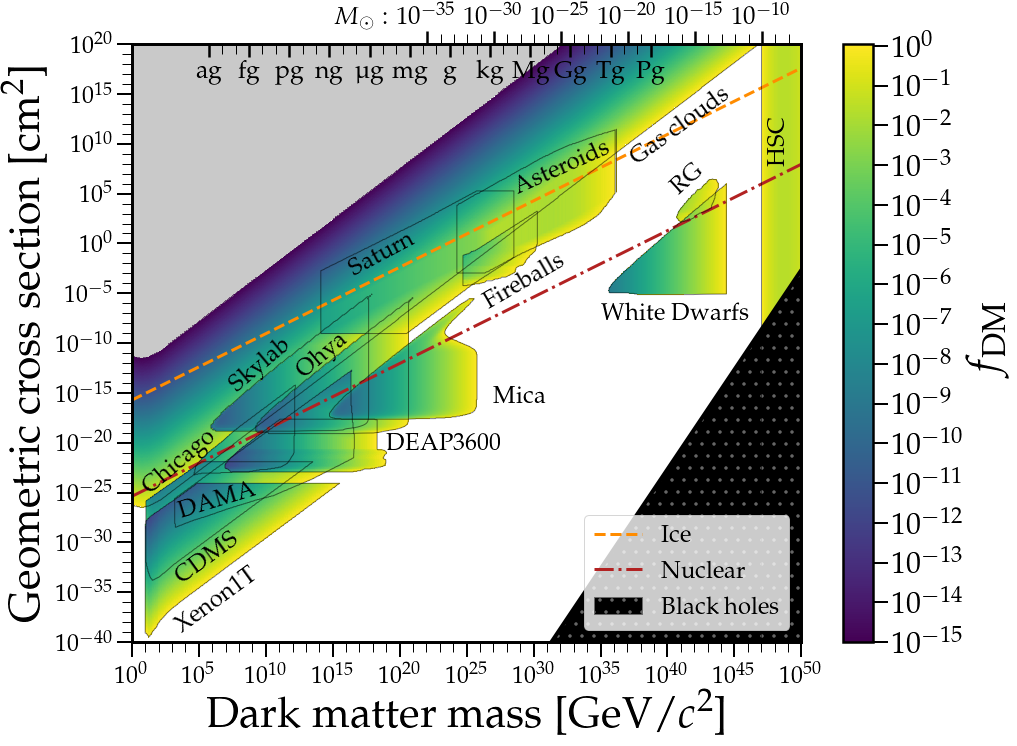}
    \caption{Constraints on macroscopic dark matter as a function of its fraction $\fdm$ of the total dark matter energy density. White space indicates that the constraints merely satisfy some $\fdm>1$. The gas cloud bounds extend into the grey region, but we cut off the heatmap scale at $\fdm=10^{-15}$. The densities of ice, nuclear matter, and black holes are included for reference. Each of the limits are detailed in the main text.}\label{fig:constraints}
\end{figure*}

\noindent The constraints collected here are not intended to represent a complete survey of the literature. Rather, we select constraints based on a combination of their competitiveness, robustness, and ease of incorporation into the extended mass distribution formalism of Sec.~\ref{sec:EMF}. There are many more potential, alternative, or future constraints on (or observations of) macroscopic dark matter, including those in the following incomplete list: large-volume liquid-scintillator neutrino detectors~\cite{Bramante:2018tos}, fluoresence detectors~\cite{SinghSidhu:2018oqs}, seismic searches~\cite{Herrin:2005kb,Cyncynates:2016rij}, sky surveys of meteors~\cite{dhakal_new_2023}, macro-macro binary formation~\cite{Diamond:2021dth}, `straight' lightning strikes~\cite{starkman_straight_2021,Cooray:2021dvp,Starkman:2022kft}, human deaths and injury~\cite{sidhu_death_2020}, the X-Ray Quantum Calorimeter~\cite{bhoonah_detecting_2021}, dark matter-induced baryon feedback from an increased supernovae rate in galaxies~\cite{acevedo_dark_2024}, the Large High Altitude Air Shower Observatory~\cite{LHAASO:2024upb}, next-generation gravitational wave observatories~\cite{Miller:2025yyx,Jiang:2025xln}, the dedicated Dark Matter and Interstellar Meteoroid Study~\cite{Kajino:2023wwa}, observable consequences of compact dark cores in stars~\cite{Bellinger:2025hrg}, underwater acoustic detectors~\cite{Cleaver:2025etu}, sporadic beam losses at the Large Hadron Collider~\cite{Liang:2026tjs}, and impulses on magnetically levitated particles~\cite{uitenbroek_first_2026}.

We assume throughout that the dark matter is interacting with Standard Model materials via elastic scattering with geometric cross-sections. In principle, this assumption may not always hold---for example, if the dark matter has internal degrees of freedom which could be excited, or interacts through a long-range mediator, then many of these limits may need recomputation. In addition, we assume that the dark matter stays tightly bound and is not broken up by any of the interactions considered; this could be violated when the binding energy of the macro becomes smaller than the energy transfer. More naively, we might be concerned if the macro is less dense than the object it is interacting with. While we do not consider the possibility here, loosely bound composite objects~\cite{bleau_secluded_2025} do have a wide range of exotic phenomenology, including multi-scattering across one or more direct detection experiments~\cite{acevedo_loosely_2025} and potential particle showers that traverse the interior of the Earth~\cite{boukhtouchen_deconstructive_2026}. Similarly, macros which can capture Standard Model particles have been studied with inelastic signatures at large-volume detectors such as IceCube~\cite{Bai:2022nsv}, with direct detection experiments, Type-Ia Supernovae, and planetary heating~\cite{Acevedo:2021kly}, and with a suite of cosmological observables~\cite{Caloni:2021bwp}.

\begin{figure*}
    \centering
    \includegraphics[width=0.8\linewidth]{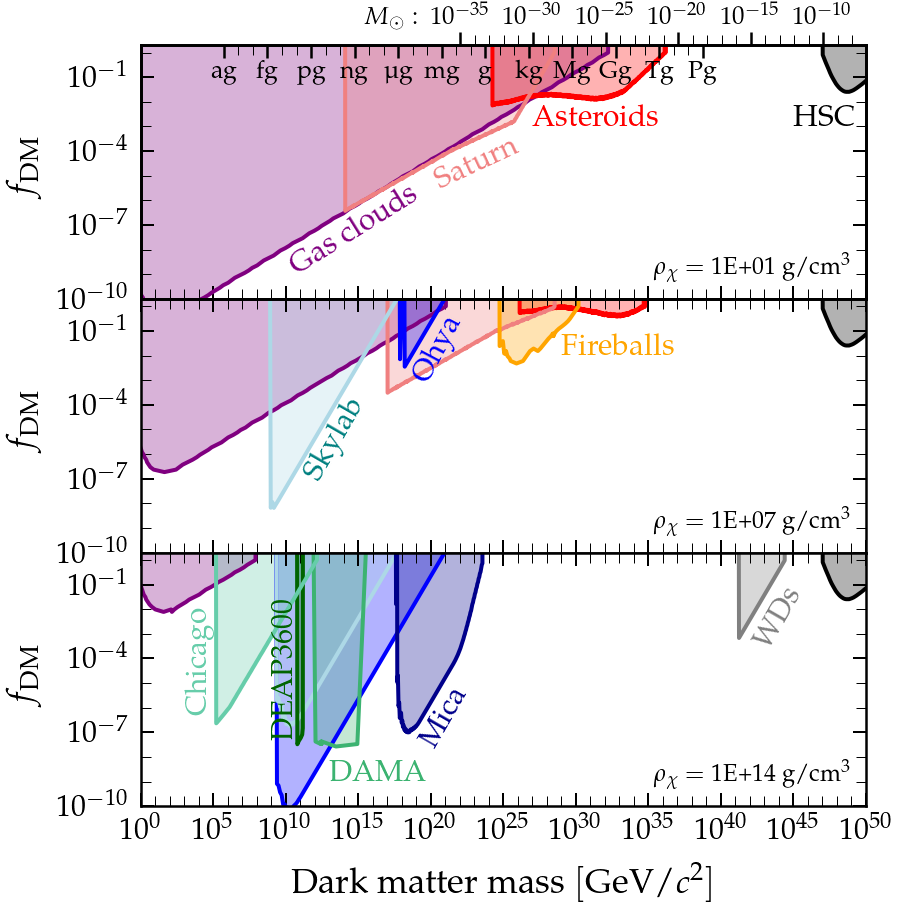}
    \caption{Constraints on macroscopic dark matter for fixed values of the internal dark matter density, as defined in Eq.~\ref{eq:sigma}. The format of this plot intentionally resembles the familiar constraints on PBHs; indeed, we find the infamously unconstrained asteroid-mass region persists even for macroscopic dark matter. The preponderance of vertical lines is due to the discrete nature of overburden or energy threshold boundaries in the constraint space.}\label{fig:contour}
\end{figure*}

The full constraints are plotted in Fig.~\ref{fig:constraints},
with the color gradient correlating to $\fdm$, the maximal allowed fraction of dark matter in macros which have a monochromatic distribution in both mass $m$ and geometric cross section $\sigma$. Plots in the $m-\fdm$ parameter space, perhaps more familiar within the PBH literature, can also be produced by fixing a mass-cross-section relation $\sigma(m)$ and computing limits along contours of this function. Specifically, in Fig.~\ref{fig:contour}, we consider a constant internal density of the dark matter $\rho_{\chi}$, at three instructive values between ice and nuclear density, and take the cross-section to be simply the geometric form $\pi R^2$ in terms of the radius $R$ of the dark matter macro.

 There are two relevant regimes which influence the overall shape of bounds. First there is the multi-scattering limit, where the dark matter is expected to scatter many times within the system. These bounds tend to take the form of an upright right triangle. The rightmost vertical boundary arises from the event rate, which scales like $1/m$. The flat horizontal bottom is typically from an energy deposition threshold: if the dark matter's cross section is too small, it will pass through the detector or material without triggering whatever phenomena is required for detection. Finally, the diagonal upper boundary is the so-called `overburden' attenuation line, where the dark matter would interact too strongly with the exterior environment, so that it arrives at the detection system without enough energy to cause observable consequences.

The second regime is the more familiar single scattering limit, relevant primarily in the low mass regime. These bounds tend to take the shape of (often, more-complicated) downward-pointing triangles. In this case, the upper-$\sigma$ boundary is set by the transition point to the multi-scattering regime, since the statistical analysis performed by these direct detection experiments is only reliable under the assumptions of single scatterings. The leftmost bound is then controlled by experimental sensitivity to the smallest single scattering momentum transfers and the corresponding highest speed flux of halo dark matter. The rising diagonal lower boundary is set by the expected number of scattering events in the detector: it scales inversely with the expected dark matter flux and hence linearly with dark matter mass. 

\subsection{Microlensing}
\noindent When a massive object passes between a source star and an observer, the starlight can be temporarily amplified as light is gravitationally focused around the lensing object, in a process known as microlensing~\cite{paczynski_gravitational_1986}. By observing many distant stars (or single stars for long periods), constraints can then be placed on the population of lensing objects. 

Our range of interest here only marginally intersects with the viable region for existing microlensing probes~\cite{kavanaghpbh}, since at small masses microlensing observables are suppressed by both diffraction and finite source-size effects. Typically microlensing constraints are computed specifically in the context of primordial black holes, but some work has been done on extended objects with radii approaching the Einstein radius~\cite{Croon:2020wpr}. In this regime, microlensing observables do not have completely trivial responses to the increasing size of the objects, and so constraints must be carefully recomputed. There is also a proposal to search for macro objects through stellar `dimming' effects, with searches similar to microlensing surveys \cite{Bramante:2024hbr}. However, within the regime considered in this paper, we will utilize results just from the Hyper Supreme Cam (HSC)~\cite{Sugiyama:2026kpv}, for which extended mass results have not yet been computed. In our work we just require that the macro size remain smaller than the Einstein ring, which is satisfied across our plotted region. 

Since PBH limits are typically plotted as a function of $\fdm$, and the characteristics of the model are fully specified by the black hole mass, it is trivial to extract $\fdm$ from existing constraints. Microlensing limits in particular are sensitive to $\fdm$ primarily through the impact on the flux of PBHs through the microlensing `tube'.

\subsection{Stellar interactions}
\subsubsection{White dwarf supernovae}
\noindent If macros collided with white dwarfs (WDs), the energy of the impact could lead to runaway fusion and a resulting Type Ia supernovae (SN). In Ref.~\cite{1992ApJ...396..649T} the runaway fusion scenario was demonstrated specifically for carbon-oxygen (CO) White Dwarfs, provided a sufficiently large amount of material in the WD core is heated to fusion temperatures $\sim 0.17- 0.68$ MeV.  Several works \cite{Graham_2015,Bramante_2015,Graham_2018,Janish_2019,Acevedo_2019,Fedderke_2020, Acevedo_2021, Diamond_2022,Acevedo_2022,Acevedo:2023cab} have used this idea to place limits on different DM models, including macros \cite{Graham_2018, Sidhu_2020,raj2024supernovaesuperburstsdarkmatter}, since the macros could transfer a considerable amount of energy from elastic multi-scattering in the white dwarf core. Constrains are then obtained by requiring that specific WDs of known ages have been able to survive long enough for us to observe them today.

\begin{table}[]
    \centering
    \begin{tabular}{|c|c|c|}
    \hline
        \textbf{White Dwarf} & \textbf{Mass $(M_{\odot})$}& \textbf{Age (Gyr)} \\
        \hline
        J0756-2001 & 1.0 & 7.72\\
        J2340+6902 &0.999 & 6.651\\
        J0206-0057 & 1.026 & 4.2\\
         SDSS J151924.44+595151.1 & 1.0 & 3.86\\
         SDSS J124726.27+102908.6 & 0.999& 3.507\\
         SDSS J014758.94+270020.9 &1.0& 3.396\\
         SDSS J102230.52+144646.8 &0.965 &3.39\\
         SDSS J161057.68+190309.3 & 1.0&3.36\\
         SDSS J101108.41+603026.8 & 1.0 & 3.112\\
         \hline
    \end{tabular}
    \caption{Examples of 1 $M_{\odot}$ WDs aged 3 Gyr or older}
    \label{Tab:WDlist}
\end{table}

The limits derived in this work will come from the continued existence of several $\sim 1  M_{\odot}$  WDs aged 3 Gyr or older.  The Montreal White Dwarf Database~\cite{MWDD:2017ASPC..509....3D} lists several such WDs in the Milky Way, which we present in Table \ref{Tab:WDlist}.  We focus on  $1  M_{\odot}$ WDs because WDs heavier than $1.05 M_{\odot}$ may have oxygen-neon (ONe) cores, which are significantly more difficult to detonate~\cite{ONe:2014ApJ...785...61S}.  Because CO and ONe WDs are difficult to differentiate, obtaining limits using heavier WDs is therefore unreliable. Meanwhile, WDs lighter than $1.05 M_{\odot}$ are only expected to have CO cores. 
% One could obtain limits using a heavy WD whose interior composition has been confirmed, but such objects are rare and will not be the focus of this work. 

We place limits on scenarios where 2.3 or more macros capable of destroying a 1  $M_{\odot}$ would be expected to collide with one within 3 Gyr (we chose 2.3 as the expected number of macro collisions to give a 90$\% $ certainty that the WD would have encountered at least 1 macro capable of destroying it in it's lifetime.)  These limits could be mildly strengthened by using the exact ages of some of the older WDs shown in Table \ref{Tab:WDlist}, but we conservatively use 3 Gyr as a benchmark WD age since there are several WDs this age or older.

The limits are bounded from above and below by the requirement that the macro can transfer enough energy into the WD to trigger runaway fusion. Macros above the upper edge of the limit slow down too quickly as they move through the outer layer of the WD to transfer sufficient energy into the inner parts of the star to trigger fusion.  Macros with $\sigma$ below the lower edge of the constrained region do not interact with a large enough patch of the WD at any given time to deposit sufficient energy to trigger runaway fusion. The limits are bounded on the right by the macro flux; we expect macros to collide with 1 $M_{\odot}$ WDs in the Milky Way at a rate,
\begin{equation}
\begin{split}
    \Gamma_{WD} &=\frac{\rho_{DM}}{m}\pi R_{WD}^2v_{vir} \left(\frac{v_{esc}}{v_{vir}}\right)^2\fdm\\&\approx 3.6 \times \fdm\left(\frac{m}{10^{20}\text{g}}\right)^{-1}\text{Gyr}^{-1}
\end{split}
\end{equation}
where $R_{WD}$ is the radius of a 1 $M_{\odot}$ WD, which we take to be $\sim 5700$ km, $\rho_{DM}\sim 0.3$ GeV/cm$^3$ is the local DM density, $v_{vir}\sim  10^{-3}$ is the virial velocity of DM particles in the Milky Way, and $v_{esc}=\sqrt{2GM_{WD}/R_{WD}}\approx2.3\times10^{-2}$ is the escape velocity at the surface of a 1 $M_{\odot}$ WD.  The ($v_{esc}/v_{vir})^2$ captures the Sommerfeld enhancement to the WD-macro cross section, otherwise known as gravitational focusing. Then to set limits we require $\Gamma_{WD}\geq 2.3/$(3 Gyr).

We now consider the conditions needed for a macro collision to cause a supernova. We require that the macro collisions with material in the WD core heat a trigger region of radius $\lambda_T$, defined below, up to a fusion temperature  $T_{crit}\approx0.17-0.68$ MeV.  Ref.~\cite{1992ApJ...396..649T} estimated the mass of nuclear matter that needed to be heated above $T_{crit}$ for runaway fusion to take place at different WD core densities and different choices of $T_{crit}$.  We extract $\lambda_T$ from these calculations by numerically fitting trigger masses as a function of $T_{crit}$ as shown in Figure 6. of Ref.~\cite{1992ApJ...396..649T} for a WD with a density of $2\times 10^8$g/cm$^3$ (the lowest density for which calculations were performed). For a given trigger mass $M_T$, $\lambda_T = (M_T/\frac{4}{3}\pi \rho)^{1/3}$.  $2\times 10^8$ g/cm$^3$ is higher than the density at all points within a 1 $M_{\odot}$ WD. We therefore extrapolate $\lambda_T$ for CO WDs using an approach similar to that described in Ref.\cite{Fedderke_2020}.  They find that for $\rho<2\times 10^8$ g/cm$^3$, the radius of the trigger region scales as

\begin{equation}
\lambda_T(\rho,T_{crit}) = \lambda_T(2\times 10^8\text{g/cm}^3\text{, }T_{crit})\left(\frac{\rho}{2\times 10^8 \text{g/cm}^3}\right)^{-2}\,.
\end{equation}
The energy needed to heat this region up to fusion energies is given by,
\begin{equation}
    E_T(\rho,T_{crit}) = \frac{4\pi}{3}\lambda_T^3 \rho c_p T_{crit}~,
\end{equation}
where $\rho$ is the (radially-dependent) density of WD material. The local heat capacity of the WD is given by $c_p = c_{p}^{ion}+\frac{1}{2}c_P^{e}+\frac{1}{4}c_p^{\gamma}$, where we have,
\begin{equation}
    \begin{split}
        &c_p^{ion} =\frac{5}{2\mu_a}\sum_i\frac{X_i}{A_i}\\
        &c_p^{e} = \frac{\pi}{\mu_a\mu_e}\frac{T_{crit}}{E_F}\left[1-\left(\frac{m_e}{E_F}\right)^2\right]^{-1}\\
        &c_p^{\gamma} = \frac{4\pi^4}{5\mu_a\mu_e}\left(\frac{T_{crit}}{E_F}\right)^3\left[1-\left(\frac{m_e}{E_F}\right)^2\right]^{-3/2}~,\\
    \end{split}
\end{equation}
where $i$ sums over the ion species that compose the WD, $X_i$ is the fraction of WD mass composed of $i$, $A_i$ is the atomic number of $i$, $\mu_a$ is the atomic mass unit, $\mu_e = \left(\sum_iX_iZ_i/A_i\right)^{-1}$ is the mean molecular mass per electron, $Z_i$ is the atomic number for ion $i$, and $E_F = m_e\sqrt{1+\left(\frac{3\pi^2\rho}{2m_e^3}\right)^{2/3}}$ is the local fermi energy of the electrons in the WD. Following Ref.~\cite{1992ApJ...396..649T}, for a 1 $M_{\odot}$ WD we take $X_C=X_O=0.5$.

As the macro moves through the WD with a velocity $v$ it loses energy to collisions with nuclei at the rate,
\begin{equation}
    \frac{dE}{dL} = v^2\sigma\rho~,
\end{equation}
where $v$ is the velocity of the macro. In order to deposit sufficient energy into the trigger region to initiate runaway fusion, the macro would therefore need to move with a velocity satisfying,
\begin{equation}
\label{eq:boomreq}
      v^2\geq T_{crit}c_p\text{Max}\left[1,\left(\frac{\lambda_T}{\sqrt{\sigma/\pi}}\right)^3\right]~,
\end{equation}
at some point in its trajectory through the star.  We check whether this is satisfied by numerically calculating the velocity of the macro as it moves through the WD and comparing this to Eq.~\ref{eq:boomreq} as a function of radius. The macro initially collides with the WD with kinetic energy
\begin{equation}
      KE_0 = \frac{1}{2}mv_{esc}^2~,
\end{equation}
which then evolves as 
\begin{equation}
      \frac{dKE}{dr} = v^2(r)\pi R^2\rho(r)-\frac{GM_{enc}(r)m}{r^2}~,
\end{equation}
where $M_{enc}$ is the WD mass enclosed at a given radius $r$.  The mass and density profiles of a 1 $M_{\odot}$ WD can be determined numerically using Chandresekar's white dwarf equations~\cite{chandrasekhar}.  We do not pick a specific value of $T_{crit}$ for this calculation, but simply check whether this requirement has been satisfied for any value of $T_{crit}$ for which numerical data was available in Ref.~\cite{1992ApJ...396..649T} $\sim0.17-0.68 $ MeV.  The upper and lower borders of the constraint region are calculated numerically but the edges can be reasonably approximated by 
\begin{align}
      \sigma&\lesssim 12\left(\frac{m}{10^{40}\text{GeV}}\right)^{1.15}\text{ cm}^2~\nonumber\\
      \sigma&\gtrsim 8\times 10^{-6}\text{ cm}^2~.
\end{align}
Within the above parameter space, where a single collision between a macro and a 1 $M_{\odot}$ WD can cause it to detonate, we therefore place limits on macro models by requiring that 
\begin{equation}
\begin{split}
      \fdm\leq&\frac{2.3}{3\text{ Gyr}\times \frac{\rho_{DM}}{m}\pi R_{WD}^2v_{vir} \left(\frac{v_{esc}}{v_{vir}}\right)^2}\\&\approx 3.8\times10^{-5}\times \left(\frac{m}{10^{40}\text{GeV}}\right)\,.
      \end{split}
\end{equation}

\subsubsection{Neutron stars}
\noindent A number of studies~\cite{Sidhu_2020, raj2024supernovaesuperburstsdarkmatter} have additionally placed limits on the macro abundance from collisions with neutron stars, which could create observable `superbursts' in the neutron star crust.  A superburst has been hypothesized to be caused by the rapid nuclear fusion of  densely packed carbon nuclei in a neutron star crust.  While superbursts remain objects of active study, nuclear flame propagation in the neutron star crust environment has not been studied as extensively as the WD environment, nor are there firmly established rates at which the carbon nuclei in the crust of neutron stars might replenish after being ignited. Hence, in our plots we have not included limits from macros triggering excess superbursts. Nevertheless, we think this mechanism is important to consider in macro searches, especially as superbursts and superburst environments become better understood.

\subsubsection{Red Giants}
\noindent Macro collisions may also affect red giant stars, as studied in Ref.~\cite{Dessert:2021wjx}. In the cores of red giants, runaway nuclear fusion can cause a helium flash, leading to an instantaneous drop in the stellar luminosity for 10-50 kyrs. Energy deposition from macros could also cause such flashes, creating observable effects on the overall luminosity function of a globular cluster. Ref.~\cite{Dessert:2021wjx} then places limits by studying the specific globular cluster M15, under the assumption that the dark matter density is simply the local Milky Way dark matter density. If in the future globular clusters are found to have a large local dark matter density, these limits could be improved.

Since a careful computation of the luminosity must be undertaken to place these limits, in principle it is not easy for us to plot $\fdm$ for their limits without redoing their entire analysis. However, since this constraint is effectively in the multi-scattering regime, where the limiting factor is the flux of dark matter through the stars, we can produce approximately correct limits by scaling them horizontally proportional to $1/m$, much the same as the white dwarf limits earlier. We plot the most conservative versions of these red giant constraints with the label `RG' on Fig.~\ref{fig:constraints}.

\subsubsection{Main sequence stars}
\noindent Finally, it may be possible with future, or even existing, optical, UV and X-ray telescopes to search for the impacts of macros on main sequence stars. In particular, Ref.~\cite{Das:2021drz} showed that the shock waves in stars that form when macros pass through them would lead to characteristic thermal UV emission on stellar surfaces which would be correlated with the local dark matter density. In particular, they showed the potential observable reach of ULTRASAT~\cite{Shvartzvald:2023ofi}, an upcoming wide-field UV space telescope, as well as the potential reach of dedicated Hubble Space Telescope observations. For the former, they consider K-dwarf stars within 1 kpc over a year's observing, while for the latter, one week of observing the globular cluster 47 Tucanae. Since these are prospective constraints, we do not include them in our plot at this stage, but rather mention them as potentially robust future constraints.

\subsection{Solar system interactions}
\noindent The collisions of large macros with objects in our Solar System could also lead to observable consequences, including craters on terrestrial surfaces~\cite{derocco_dark_2025}, fireballs on Earth~\cite{sidhu_macroscopic_2019,dhakal_new_2023}, or the catastrophic destruction of asteroids and planetary rings~\cite{picker_dark_2025}. 

In ref.~\cite{derocco_dark_2025}, it was found that macros puncturing the Jupiter's moon Ganymede could deposit subsurface liquid onto its surface, producing potentially visible surface features for the upcoming missions Europa Clipper and JUICE. Since these are forecasted limits, we do not yet include them here.

Ref.~\cite{sidhu_macroscopic_2019} placed macro constraints from the (non)observation by bolide camera networks of fireballs entering Earth's atmosphere. These constraints are effectively flux-limited, so to compute $\fdm$ we again simply rescale these `fireball' limits horizontally as $1/m$.

The destruction of asteroids and ring particles was studied by one of the authors in Ref.~\cite{picker_dark_2025}. Catastrophic destruction here is defined as the requirement that after the collision, the largest remaining fragment is smaller than half of its original mass, and less than half of the fragments gravitationally re-accumulate. If such collisions are expected to happen on timescales much shorter than their expected ages, we can place constraints on macro dark matter.

For asteroids, we can estimate their ages from simulations of their average collisional lifetime from impacts with other asteroids~\cite{bottkejr_linking_2005}. For Saturn's rings, we take their age to be young, at $100$ Myr~\cite{crida_age_2025}. Arguments about the stability of structures in Saturn's rings, as well as the exposure age from dust bombardment place the age in the $100-400$ Myr mark, although that may yet be an underestimate~\cite{saturn:RICERCHI2026117029}. Indeed, the dynamical age of Saturn's rings is consistent with ages in the $1.5-4.5$ Gyr range. That there is no consensus on the formation mechanism of the rings only increases this uncertainty, so we take the most conservative (i.e., young) estimate here; however, older ring ages could improve the constraints by an order of magnitude. 

We recompute the constraints of Ref.~\cite{picker_dark_2025}, including now the dark matter fraction $\fdm$. For a given mass of impacted object (asteroid or ring particle), constraints are placed by considering the region where sufficient energy is injected into the object to catastrophically destroy it, with a collision rate high enough that we would not expect any to survive until today. Details of this calculation can be found in Ref.~\cite{picker_dark_2025}.

For a given object mass, we then get constraint regions similar to the multi-scattering regime, but without an overburden line, since the objects are freely floating in space. For the upper boundary, our calculation is not reliable when the macro is radially larger than the object in question. Since asteroids and ring particles come in a wide mass spectrum, the full constraints are then found by integrating over their full mass spectra. This ultimately leads to a relatively non-trivial dependence of the bounds on $\fdm$, since the energy required for catastrophic destruction is sensitive to the material regimes of bodies of different sizes.

Here we differ slightly from Ref.~\cite{picker_dark_2025} in that we modify the event rate limit here to now require $\ln(10)$ macros to collide with the body over its lifetime in order to place constraints, making the limits more consistent with the 90\% confidence intervals used by most other limits collected here. Limits from the survival of both S-type asteroids and Saturn's dense A ring are included on Fig.~\ref{fig:constraints}.

\subsection{Single- and multi-scattering direct detection}

\subsubsection{Multi-scattering track detectors}
\noindent We begin our discussion of detector-based heavy dark matter searches with the Skylab~\cite{Bhoonah:2020fys}, Ohya~\cite{Bhoonah:2020fys}, and Mica~\cite{micadm:PhysRevLett.74.4133,jacobs_macro_2015,Acevedo:2021tbl,Boukhtouchen:2026rfz} bounds on macros. Skylab was a search for high-Z cosmic rays aboard the Skylab satellite~\cite{skylab:1978ApJ...220..719S}, and Ohya was a search for magnetic monopoles in a quarry north of Tokyo~\cite{ohya:PhysRevLett.66.1951}; both searched for tracks in planar plastic etch detectors. Mica was also a search for magnetic monopoles, looking for damage tracks in mica sheets~\cite{mica:PhysRevLett.56.1226}. All three have been recast into multi-scattering dark matter bounds. We present a unified treatment to compute $\fdm$ for all three of these bounds. X-ray-based readout of mica, along with quartz- and olivine-based searches have also been proposed as potential future paleodetectors for macros~\cite{Ebadi:2021cte,Baum:2023cct,Boukhtouchen:2026rfz}.

As dark matter passes through the detector, it must maintain enough energy throughout to trigger the detector. This threshold energy is most conveniently stated as a cross-section dependent threshold speed $v_{\mathrm{min}} \propto R^{-1}$. The value of $v_{\mathrm{min}}$ at a reference dark matter radius of $R=1~$nm for each experiment is given in Table \ref{tab:paleo}, along with other experimental parameters given in this section. Dark matter must pass through the overburden (either the Earth or the exterior of the Skylab satellite) and the entirety of the detector while remaining above this threshold speed. The speed loss from traversing the overburden and detector is exponential in the total column density $X$ across the dark matter's path,
\begin{equation}\label{eq:elossmimp)}
  v_f =  v_0e^{-\sigma X/m}.
\end{equation}

As discussed in Ref.~\cite{Boukhtouchen:2026rfz}, there are two refinements that should be made to the mica constraints. Firstly, the threshold speed used in older analyses was too small; damage to a sheet of mica from dark matter passing through at this speed would may have annealed away over the geological lifespan of the mica, given results indicating rapid annealing of alpha tracks in micas \cite{yuan2009annealing}. Instead, we use a higher threshold energy deposition, with corresponding speed value in Table~\ref{tab:paleo} which is matched to, and can be validated by, the observation of fission tracks in mica from trace radioactive elements. Secondly, the original monopole search in mica discarded any mica slab with visible damage, which we take to be any damage track with a radius greater than $10~\mu$m. This can happen if the dark matter passing through the mica has a speed greater than $1.42\times10^{4}\ \dfrac{\text{km}}{\text{s}}\left(\dfrac{1~\text{nm}}{R}\right)$ thereby melting an optically-resolvable vitrified track; indeed, if the macro has a radius greater than $10~\mu$m, this can also lead to a hole apparent to the naked eye. Hence the bounds shown here do not exclude parameter space where the monopole search of \cite{mica:PhysRevLett.56.1226} would have discarded the sample or sample area, as visibly damaged, and tracks producing such damage are not counted in our exclusion.

The expected number of tracks over the lifetime of a planar detector is given by the following integral over dark matter velocity-space and over time,
\begin{align}\label{eq:Nmimps}
N_{\mathrm{exp}}(m,\sigma) = {} & A\, \rho_{\mathrm{DM}}/m \int_0^{T} \!dt
\int_{v_{\mathrm{min}}^{*}}^{v_{\mathrm{max}}^{*}} \!\!dv\, v^2 \notag \\
& \times \int \! d\hat{v}\; f(-\mathbf{v} \mid t)\,
\big[\mathbf{v} \cdot \hat{n}(t)\big]\,
\Theta\big(\theta_{\mathrm{acc}} - \theta_D\big)
\end{align}
where $\hat{n}(t)$ is the normal vector of the detector over time modelled as described in Table \ref{tab:paleo}, $\theta_D=\arccos(\hat{v}_D \cdot\hat{n}(t))$ is the angle the dark matter trajectory makes with the detector's normal and $v^{*}=ve^{\sigma X/m}$ accounts for the exponential loss of speed through the overburden. All dark matter is assumed to pass through the same amount of overburden, conservatively taken to be the maximum possible. Additionally, Eq. \ref{eq:Nmimps} uses the surface area $A$ and lifetime $T$ of the detector, as well as the `acceptance angle' of the detector $\theta_{\mathrm{acc}}$, which is the maximum angle relative to the detector normal the dark matter must hit to leave a signal. 
Next, $f(\mathbf{v} \mid t)$ is the dark matter velocity distribution at time $t$. The mica bound neglects the astrophysical evolution of the dark matter halo over geological timescales and assumes an unchanging Standard Halo Model with standard parameters. On the other hand, the most recent Skylab and Ohya bounds use a more complex estimate for the local halo obtained from simulations, particularly by including the effect of the Large Magellanic Cloud~\cite{Bozorgnia:2025lsl}

The value for $\fdm$ at a particular value of mass and cross section is then $\fdm=\frac{N_{\mathrm{obs}}}{N_{\mathrm{exp}}(m,\sigma)}$, where $N_{\mathrm{obs}}$ is the number of observed dark-matter candidate signals. All three experiments made no positive detections, so $N_{\mathrm{obs}}$ is taken to be $-\ln(1-0.9)\approx2.3$ as a 90\% confidence value, assuming Poisson statistics. From Fig. \ref{fig:constraints}, the contours of constant $\fdm$ appear to lie at constant mass, suggesting that the constraint on $\fdm$ is mainly determined by rescaling the total flux through the detector.

\begin{table*}[t]
\centering
\renewcommand{\arraystretch}{1.4}
\setlength{\tabcolsep}{6pt}
\newcommand{\ccell}[1]{\parbox[c]{3.2cm}{\centering #1}}
\newcommand{\ccellwide}[1]{\parbox[c]{6.7cm}{\centering #1}}
\newcommand{\Tstrut}{\rule{0pt}{2.8ex}}
\newcommand{\Bstrut}{\rule[-1.2ex]{0pt}{0pt}}
\begin{tabular}{|l|c|c|c|}
\hline
\textbf{Quantity} & \textbf{Skylab} & \textbf{Ohya} & \textbf{Mica} \\
\hline
Detector Area (m$^2$) & \ccell{$1.17$} & \ccell{$2442$} & \ccell{$0.24$} \\
\hline
Detector Lifetime (yr) & \ccell{$0.70$} & \ccell{$2.1$} & \ccell{$5\times10^{8}$} \\
\hline
Acceptance Angle & \ccell{$60^{\circ}$} & \ccell{$18.4^{\circ}$} & \ccell{$60^{\circ}$} \\
\hline
Overburden (g\,cm$^{-2}$) & \ccell{$4.0$} & \ccell{$1.15\times10^{4}$} & \ccell{$1.07\times10^{7}$} \\
\hline
Lower Threshold Speed at $R{=}1$\,nm (km/s) & \ccell{$1.46$} & \ccell{$1.08$} & \ccell{$8.85$} \\
\hline
Upper Threshold Speed at $R{=}1$\,nm (km/s) & \multicolumn{2}{c|}{\ccellwide{N/A}} & \ccell{$1.42\times10^{4}$} \\
\hline
Upper Cutoff Cross Section (cm$^2$) & \multicolumn{2}{c|}{\ccellwide{N/A}} & \ccell{$3.14\times10^{-6}$} \\
\hline
\rule{0pt}{14pt}
Model of Detector Orientation & \ccell{Random orientation over time} & \ccell{Flush with Earth's surface at $36.6^{\circ}$N} & \ccell{Random orientation over time} \\[6pt]
\hline
Model of Local Velocity Distribution & \multicolumn{2}{c|}{\ccellwide{\Tstrut LMC + Milky Way Simulation\\(see \cite{Bozorgnia:2025lsl} for details)\Bstrut}} & \ccell{\Tstrut Standard Halo Model\\($v_0{=}222$, $v_E{=}232$,\\$v_{\rm esc}{=}503$ km/s)\Bstrut} \\
\hline
\end{tabular}
\caption{The differing experimental and model parameters for the Skylab, Ohya, and Mica constraints}
\label{tab:paleo}
\end{table*}

\subsubsection{Other direct detection experiments}

\noindent The `Chicago' bound was derived in Ref~\cite{cappiello_new_2021}, wherein two separated liquid scintillators were used to look for coincident signals from a macro trajectory at the University of Chicago~\cite{Collar:2018ydf,Blanco:2019lrf}. Similar to the other multi-scattering bounds, we estimate the $\fdm$ dependence of both these bounds by simply rescaling the bounds according to the diminished flux expected at smaller $\fdm$, which should be appropriate in this regime.
% Many multiscatter searches are threshold limited ($i.e.$ they require a minimum cross-section to trigger the detector), and so for these experiments the $\fdm$ rescaling results in the bound rescaling to lower masses, horizontally in the (mass, cross section) plane.

A number of more traditional direct detection dark matter experiments have also performed searches relevant for the heavy dark matter parameter space, including single-scatter and multi-scatter searches, where in a multi-scatter search the heavy dark matter is expected to leave recoils along a straight track through the detector \cite{bramante_saturated_2018,Bramante:2018tos}. These direct detection searches include single- and multiscatter analyses at XENON1T, LZ, DAMA, and CDEX-10 \cite{XENON:2023iku,LZ:2024psa,Bernabei:1999ui,CDEX:2025gtl}. The most trenchant heavy dark matter search to date was completed by DEAP3600~\cite{DEAP-3600:2017ker}, which has the advantage of having the largest surface area of any active direct detection experiment~\cite{DEAPCollaboration:2021raj}. 

At the lower end of the mass range, single-scatter searches tend to be more relevant, and can also be used to place constraints on dark matter with a large cross-section. Constraints from XENON1T~\cite{XENON:2018voc,Clark:2020mna}, CDMS~\cite{CDMS:2000lgz,CDMS:2002moo,Kavanagh:2017cru}, and DAMA~\cite{Bernabei:1999ui} were compiled in Ref~\cite{bhoonah_detecting_2021}, which we use as the basis for our estimation. Because these bounds are single-scatter experiments, we estimate their $\fdm$ dependence by rescaling according to the diminution in flux at smaller $\fdm$, corresponding to a lower number of expected scatters in the detectors. The resulting contours of constant $\fdm$ lie at constant $\frac{\sigma}{m}$. %results in a decreased sensitivity in cross-section, and a corresponding vertical rescaling of bounds in the (mass, cross section) plane with decreasing $\fdm$. \zac{explain this better?} 

\subsection{Cosmology, large-scale astrophysics, and galactic gas clouds}
\noindent A number of constraints on dark matter-nucleon scattering have been computed for cosmological and astrophysical structures. Typically, such constraints provide a diagonal `ceiling' which limits the maximum size of dark matter macros. These include constraints from a dark matter-baryon drag force which would affect Planck and Lyman-$\alpha$ observations~\cite{Dvorkin:2013cea,Gluscevic:2017ywp}, and from the potential washing-out of small scale structure observed as Milky Way satellite galaxies~\cite{Nadler:2019zrb,DES:2020fxi}.
%(We note that in some works, e.g.~Refs.~\cite{Acevedo:2021kly,Caloni:2021bwp}, macro limits have been considered for particular models under assumptions about the capture of baryons onto the macros themselves.)
However, none of these constraints were obtained with the intention of extending their limits to arbitrarily high masses, and their analyses have not been done in the macro regime. Nevertheless, it is possible to estimate the validity of the cosmological bounds \cite{Dvorkin:2013cea,Gluscevic:2017ywp,Nadler:2019zrb,DES:2020fxi} in the macro regime, and specifically up to what mass we expect these bounds remain valid. 

In the heavy macro regime, the Boltzmann equations employed in cosmological studies are influenced by the presence of macros through a drag term that depends on $\fdm \times \sigma/m$; the crucial question is, up to what mass should we expect this fluid approximation to be valid? A few estimates indicate that the validity of cosmological bounds extends beyond the reach of our plot in Figure \ref{fig:constraints}. First, let us assume that the macros have formed into their massive states before the cosmological epochs considered hereafter. Then we can determine whether enough macros populate the relevant comoving cosmological scales during the generation of perturbations that they may be treated as a fluid, i.e. that many macros exist in each comoving sphere of size $k^{-1}$. The cosmological fluid cell corresponding to a perturbation with comoving wavenumber $k$ would be affected by macros contained in a comoving ball of radius $k^{-1}$. The number of macros in this ball is $N = (4\pi/3)\fdm\bar\rho_{\rm DM}k^{-3}/m$, here given in terms of present day average dark matter density $\bar\rho_{\rm DM}\simeq1.3\times10^{-6}$ GeV/cm$^3$, where we have taken cosmological inputs from \cite{Planck:2018vyg}. Since dark matter (and macro number) is conserved in comoving volumes, $N$ may be evaluated with present-day quantities, and so requiring $N \gg 1$ leads to
\begin{align}\label{eq:granularity}
    m \ll \frac{4\pi}{3}\frac{\fdm \bar\rho_{\rm DM}}{k^{3}} \simeq 1.6 \times10^{68}\text{ GeV}~\fdm\left(\frac{k}{\text{Mpc}^{-1}}\right)^{-3};
\end{align}
this is also just the dark matter mass inside the horizon when the $k$ mode enters, at $1+z_k = ck/(H_0\Omega_r^{1/2})\simeq4.6\times10^{5}\,(k/\text{Mpc}^{-1})$, where this relation holds for a radiation-dominated cosmological epoch. Eq.~\eqref{eq:granularity} clearly does not restrict the bounds shown in Figure \ref{fig:constraints} for the Mpc scales relevant for CMB and Lyman-$\alpha$ bounds. Using similar arguments, we find that the macro parameter space we consider is also comfortably pointlike on horizon entry scales relevant for the CMB and Lyman-$\alpha$ constraints, with radii $R=\sqrt{\sigma/\pi}\lesssim10^{9}\text{ cm}$, which is much smaller than comoving scales relevant for cosmological bounds on macros. Nevertheless, partly out of an abundance of caution, and partly because the bound we will presently discuss covers essentially the same parameter space, in Figure \ref{fig:constraints} we have not shown bounds on macros from cosmological analysis of baryonic scattering---it will be interesting to revisit this topic in future work.

As just mentioned, a constraint similar to the cosmological one has been applied to high mass macros from the heating of galactic gas clouds from dark matter interactions~\cite{Chivukula:1989cc}. In Ref.~\cite{bhoonah_detecting_2021}, the gas cloud G357.8-4.7-55~\cite{mcclure-griffiths_atomic_2013} constrained macros by requiring that the dark matter heating rate not exceed the volumetric cooling rate today \cite{Bhoonah:2018gjb,Bhoonah:2018wmw}. This constraint is competitive with both the CMB and Milky Way satellite constraints, and has been computed for geometric contact interactions to high masses, so we include it as our primary ceiling constraint, plotted in Fig.~\ref{fig:constraints}. Since the heating effect scales with $\sigma/m$, the scaling of the bound with $\fdm$ follows the same form as for low mass direct detection experiments. At sufficiently small $\fdm$ we cut off the constraint here (to avoid visually obscuring $\fdm$ for the other constraints) and fill the remaining region in grey. The maximal allowed $\fdm$ indeed grows without bound along this diagonal, until the macros become so rare that the heating is no longer continuous and volumetric. The expected number of macros inside the cloud at any moment is $N_c = \fdm \rho_{\rm DM}V_c/m$, where $V_c=(4\pi/3)R_c^3$ is the approximate cloud volume and $\rho_{\rm DM}$ the ambient dark matter density at its location. Requiring that macros always be present in the cloud to produce the heating effect sets an upper limit on the constrainable macro mass of 
\begin{align}\label{eq:occupancy}
    m &\lesssim \fdm \rho_{\rm DM}\frac{4\pi}{3}R_c^3 \nonumber\\
    &\simeq 6\times10^{59}\text{ GeV}~\fdm\left(\frac{\rho_{\rm DM}}{5\text{ GeV/cm}^3}\right)\left(\frac{R_c}{10\text{ pc}}\right)^{3},
\end{align}
which lies outside parameter space restricted in Figure \ref{fig:constraints}. A weaker cutoff on the validity of the mass regime would require a few macro transits across the cloud ($t_{\rm cross}$) per cloud cooling time $t_{\rm cool}$; this would relax Eq.~\eqref{eq:occupancy} by a factor $\sim (3/4)\,t_{\rm cool}/t_{\rm cross}\sim20$, but would require detailed calculation of conduction times for spreading the locally deposited track energy, so here we quote the more conservative occupancy condition. Finally we note that both for the cosmological bounds, and the gas cloud heating bounds, some of the grey region may be unrealistic from a model-building standpoint: it is reasonable to ask whether a ``geometric" cross-section of $10^{20}~{\rm cm^2}$ (planet size) is obtainable for a GeV mass state that couples to the Standard Model; we leave exploration of this question to future work.

\section{Extended mass functions}\label{sec:EMF}

\noindent If macroscopic dark matter is some sort of composite object, then it must have been assembled in the early universe. In general, these kind of processes lead to characteristic extended mass functions. Since the tails of these distributions can have wide extent, it is not completely straightforward to determine the viability of these models purely from the monochromatic constraints in Fig.~\ref{fig:constraints}. Of particular consequence would be the scenario where the macros do not contain the totality of their constituent particles, and some remnant population of either uncoupled, or very small bound states, persists in the universe.

\begin{figure*}
\begin{subfigure}{\textwidth}
    \centering
    \includegraphics[width=.9\linewidth]{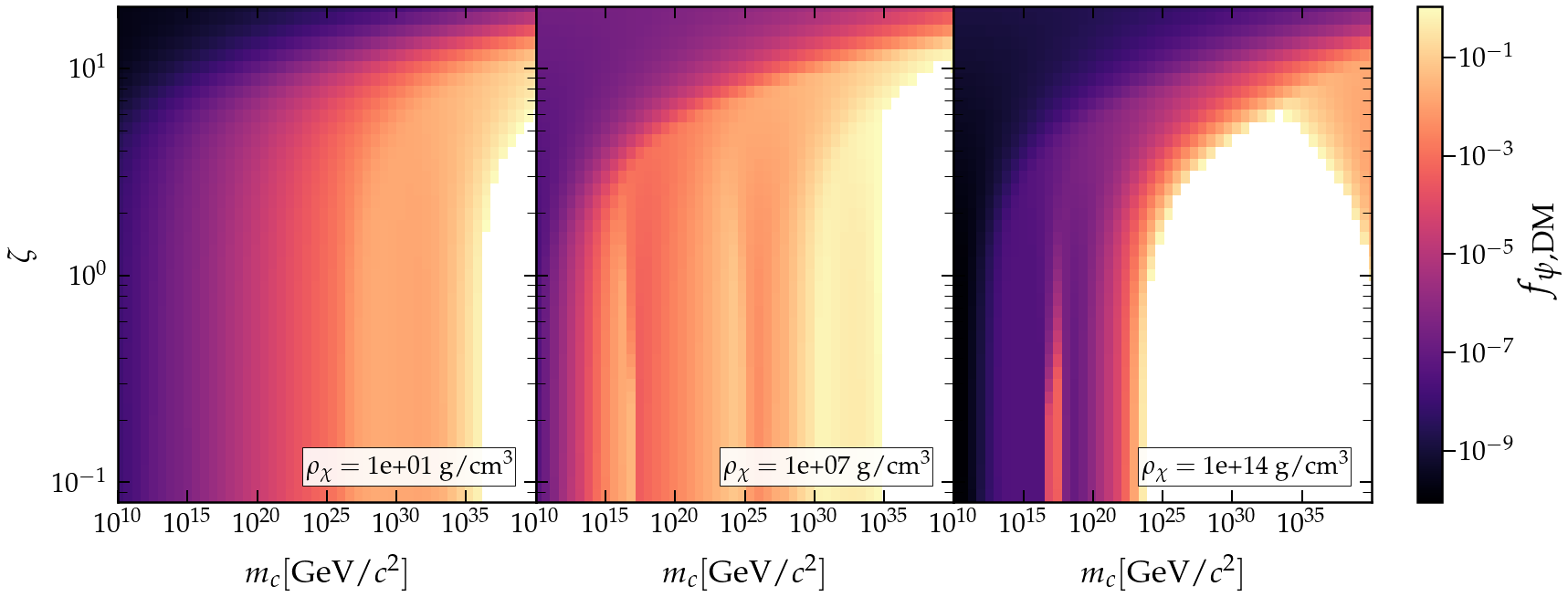}
    \caption{Constraints for a lognormal mass function, for three choices of constant macro density, in terms of the peak mass $m_c$ and log-width $\zeta$. We constrain our search space here somewhat because a very wide distribution near the edges of the search space would require us to include constraints beyond the boundaries of Fig.~\ref{fig:constraints}}\label{fig:lognormal}.
\end{subfigure}
\begin{subfigure}{\textwidth}
    \centering
    \includegraphics[width=0.9\linewidth]{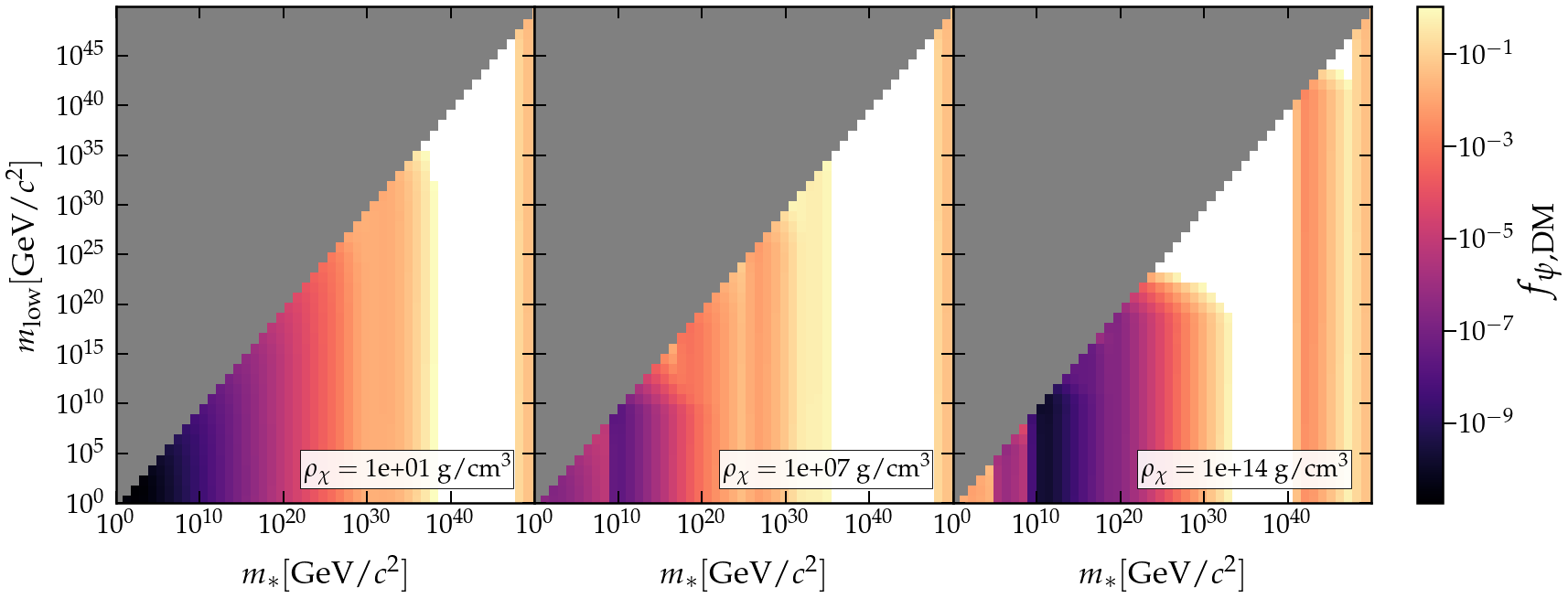}
    \caption{Constraints on a Press-Schechter mass function with characteristic mass $m_*$ and lower mass limit $m_{\rm low}$, for three choices of macro density. The grey region covers the unavailable area where the lower mass would be more massive than the heavy mass cutoff.}\label{fig:ps}
\end{subfigure}
\begin{subfigure}{\textwidth}
    \centering
    \includegraphics[width=0.9\linewidth]{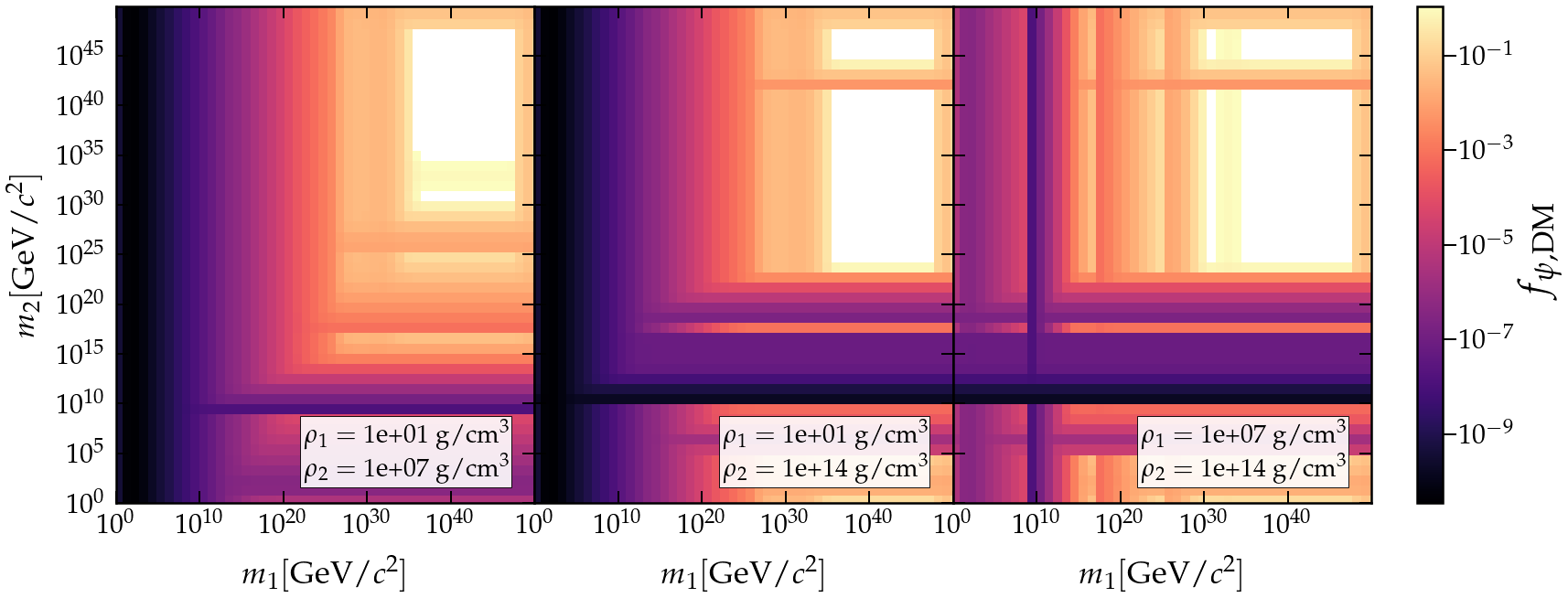}
    \caption{Constraints on a bimodal mass function for macros for $f_{12}=0.5$, at masses $m_1$ and $m_2$. This time, we only show the cases where each of the macros are at differing internal densities.}\label{fig:bimodal}
\end{subfigure}
\end{figure*}

To derive maximally accurate constraints for extended mass functions, one would need to recompute all constraints from scratch, since physical observables in principle depend nontrivially on the mass and radii of the dark matter. In the absence of the original analysis pipelines, however, there is still a formalism which allows us to approximate constraints for any given extended mass function using our monochromatic $\fdm$ constraints. 

In order to compute constraints for extended mass functions, we first need to assume that the macros have some mass-cross section relation $\sigma(m)$. Then, following Refs.~\cite{carr_primordial_2017,bellomo_primordial_2018}, we define the extended mass function in the following way:
\begin{align}
    \psi(m) &\propto m \frac{\dd n}{\dd m},\nonumber\\
    1 &= \int\dd m~ \psi(m)~,
\end{align}
where $\dd n/\dd m$ is the number density of macros in the range $[m, m+\dd m]$ and the second line sets the normalization of the distribution. We also define $\fpdm$, the maximal allowed fraction of the dark matter in macros described by the mass function $\psi$. It can be derived by integrating each constraint on our monochromatic constraint plot along the contour defined by the mass relation $\sigma(m)$, and then adding in quadrature:
\begin{align}
    \frac{1}{\fpdm^{2}} = \sum_i \left( \int_{\sigma} \dd m \frac{\psi(m)}{\fdm(m,\sigma(m))}\right)^2~,
\end{align}
where $i$ iterates over all constraints and $\fdm(m,\sigma)$ is the value of the monochromatic dark matter fraction constraint. The constraints are added in quadrature following the procedure for finding the $\chi^2$ statistic for multiple independent observables, as computed in the Appendix of Ref.~\cite{carr_primordial_2017}.

Note that the above prescription is only valid when all of the considered constraints are \textit{linear} in the dark matter number density. If, for example, a limit depended on collisions between macros, the constraint would go with the square of the dark matter density, rather than linearly. Luckily, all of our considered constraints are effectively linear in the macro density, so we can proceed.

Let us return briefly to the cross-section relation. A simple and motivated choice of mass-cross-section relation might be that of constant density matter, where we have,
\begin{align}\label{eq:sigma}
    \sigma(m) = \pi\left(\frac{3}{4\pi}\frac{m}{\rho_\chi}\right)^{2/3}~,
\end{align}
in terms of some constant internal dark matter density $\rho_\chi$, and where we are assuming that the interaction cross-section scales geometrically as $\pi R^2$. In fig.~\ref{fig:contour} we plot the macro constraints along such contours, and will take this constant density cross-section relation for the remainder of this paper. In practice, however, our code is structured so that these plots can be produced for any mass-cross-section relation.

For physical motivation, we might consider fermionic nuggets in an asymmetric dark sector~\cite{gresham_nuclear_2017,lu_black_2025}. Here constant density would be an acceptable mass-radius relation in the limit where the nuggets are sufficiently large such that they are saturated, interacting with roughly geometric cross-section. When the nuggets are too small, however, the constant density relation fails, and they enter regimes where they satisfy $R\sim N^{2/3}$ and $R\sim N^{-1/3}$, in terms of the number of constitutents in the composite $N$. Furthermore, one could not simply approximate the SM interaction cross-section as $\pi R^2$ at these scales. In this case, one would need to take a different cross-section relation. We mention this in order to demonstrate the need for a generic function $\sigma(m)$, but will proceed with the simpler Eq.~\ref{eq:sigma} from here. 

\subsection{Lognormal distribution}
\noindent For primordial black holes, the lognormal distribution is a well-studied extended mass function, given by:
\begin{align}
    \psi_{\rm LN}(m) \equiv \frac{1}{\sqrt{2\pi}\zeta m}\exp\left(-\frac{\ln^2(m/m_c)}{2\zeta^2}\right)~,
\end{align}
where $\zeta$ here is the width of the function and $m_c$ is the position of the peak. In Ref.~\cite{gresham_early_2018}, fermionic nuggets which are constructed with a dark-nucleosynthesis pipeline resulted in a qualitatively similar mass function, so we compute limits for the lognormal mass function here as a useful benchmark.

In Fig.~\ref{fig:lognormal} we compute the constraints on a lognormal distribution, taking again the above cross-section relation for regular matter at three values of density. It is not surprising to find that wider distributions do not allow one to evade constraints; rather, the tails of the distribution tend to strongly constrain such a population.

\subsection{Press-Schechter distribution}
\noindent Another physically motivated mass function might be the Press-Schechter distribution. We might expect such a distribution when we have a dark sector with an attractive force stronger than gravity, causing structure formation in the dark sector well before matter-radiation equality~\cite{Gradwohl:1992ue,gubser_structure_2004,nusser_structure_2005,amendola_primordial_2018,savastano_primordial_2019,flores_primordial_2021,domenech_cosmology_2021,flores_gravitational_2022,flores_inhomogeneous_2022,flores_structure_2023,lu_black_2025,domenech_halo_2023}. This would result in a distribution of small-scale dark halos. Such halos can merge with other and collapse, since the mediator of the long-range force can additionally allow for efficient energy dissipation. The collapse of the halos can then lead either to PBHs~\cite{flores_primordial_2021} or to Fermi balls~\cite{lu_black_2025}, so their mass function may have the Press-Schechter distribution imprinted on it:
\begin{align}
    \psi_{\rm PS} \propto \frac{1}{\sqrt{\pi}~m}\left(\frac{m}{m_*}\right)^{1/2}\exp\left(-\frac{m}{m_*}\right)~,
\end{align}
where $m_*$ is the characteristic upper mass limiting structure formation, imposed by some physical consideration related to the mechanics of the structure formation. While the Press-Schechter distribution above extends to arbitrarily low masses, one also usually imposes a low-mass cutoff related to the physics of structure formation. In the usual case of structure formation, the Jean's length would comprise an appropriate cutoff; for the dark sector case, it may be more correct to use the dark matter free-streaming length as the scale at which structures are washed out. Either way, we introduce a second scale $m_{\rm low}$, below which the distribution drops to zero. After including this cutoff, the full distribution then must be correctly normalized, which we impose numerically. We show the constraints for this distribution as a function of both mass scales in Fig.~\ref{fig:ps}.

\subsection{Bimodal distribution}
\noindent Finally, it may be the case that one has two or more independent distributions to be combined. We demonstrate such a case here with a simple bimodal distribution, where we have two macro masses $m_1$ and $m_2$, and a fraction $f_{12}$ which gives their relative contributions to the total dark matter energy density, $\rho_{m_1}\equiv f_{12}\rho_{\rm DM},~\rho_{m_2}\equiv (1-f_{12})\rho_{\rm DM}$. Moreover, we take each of the two to have different internal densities. In practice it is easier to take a narrow lognormal peak for each as an approximation of a delta function, since our code is already optimized for arbitrary distributions.

We plot these constraints in Fig.~\ref{fig:bimodal}, keeping $f_{12}=0.5$ fixed, and showing only the three cases where the two components have different densities, since the alternative can be trivially read off the existing monochromatic plots in Fig.~\ref{fig:contour}.

\section{Conclusions}
\noindent In this paper we compiled, and in some cases recomputed, the constraints on macroscopic dark matter, retaining for the first time their dependence on $\fdm$, the fraction of the total dark matter in macros at a given mass $m$. We then showed how this could be mapped to constraints on extended mass functions for specific macro candidates. Such constraints are generically required for macroscopic dark matter candidates, since the assembly processes which are able to build such heavy, composite candidates often result in extended mass distributions. Moreover, if some subcomponent of the dark matter does not end up in large composite objects, it may be strongly constrained. Indeed, even a small part of the tail of these extended distributions can lead to severe constraints for many scenarios of macroscopic dark matter formation. For demonstration, we explicitly computed three examples; the lognormal distribution, a generic Press-Schechter distribution, and a bimodal scenario with differing internal densities.

The code used to compute and plot these constraints has been made public at Ref.~\cite{macrolimits}, with more information in App.~\ref{sec:app} on how to use it. We also humbly request that anyone computing dark matter constraints (especially when complicated statistical treatments are required) please explicitly compute the dependence on $\fdm$, so that constraints can be added to the public repository as needed.

\section*{Acknowledgements}
\noindent We thank Ciaran O'Hare for gratuitous use of both code and limits compiled in his DirectDetectionPlots repository~\cite{oharedd}. We thank Will DeRocco and Glenn Starkman for helpful conversation. This work made use of N\textsc{um}P\textsc{y}~\cite{numpy2020Natur.585..357H}, S\textsc{ci}P\textsc{y}~\cite{scipy2020NatMe..17..261V}, and M\textsc{atplotlib}~\cite{mpl4160265}. We are grateful to the Mainz Institute for Theoretical Physics (MITP) of the Cluster of Excellence PRISMA+ (Project ID 390831469) for hospitality and support of ZSCP during a portion of the completion of this paper.  This research was undertaken thanks in part to funding from the Natural Sciences and Engineering Research Council of Canada through the Arthur B. McDonald Canadian Astroparticle Physics Research Institute.

\appendix
\section{Code}\label{sec:app}
\noindent The code for generating all the figures in this paper is hosted on GitHub at Ref.~\cite{macrolimits}, and functions as a class in Python called \verb|macro|. We briefly summarize some of the more useful methods in the class here. Required packages are N\textsc{um}P\textsc{y}~\cite{numpy2020Natur.585..357H}, S\textsc{ci}P\textsc{y}~\cite{scipy2020NatMe..17..261V}, and M\textsc{atplotlib}~\cite{mpl4160265}.

\subsection{Individual constraint methods}
\noindent For individual constraints, the $\fdm$ data is stored as two-dimensional NumPy arrays in the \verb|/macro_constraints/| directory, where each element gives the value of $\fdm$, and the two axes represent the mass and cross-section. The actual values of these axes are different for each respective constraint, and are most easily called by invoking the relevant method for any individual constraint; e.g., \verb|macro.asteroid(original=False)|, or \verb|macro.skylab(original=False)|. If \verb|original==True|, each method will return the mass axis, the cross-section axis, and the two-dimensional $\fdm$ array. Otherwise, the method will return just the $\fdm$ array, resized to the attributes \verb|macro.marray| and \verb|macro.sarray|, which are defined at initialization and set the constraint plot axes.

\subsection{Constraint plot}
\noindent The constraint plot shown in Fig.~\ref{fig:constraints} can be produced by calling \verb|macro.plot_constraints()|, with the arguments \verb|lines=True| and \verb|text=True| respectively determining whether or not individual constraints are outlined and text labels are included on the figure. To include additional constraints (or to remove them), the constraint data should be first be called with a method like the above, and then included  in (or removed from) \verb|alist|.

\subsection{Contour constraints}
\noindent To modify the mass-cross-section relation, the method \verb|macro.cross_section_func(m,param)| should be appropriately modified. It takes as an argument an array of dark matter masses and a list of parameters, returning an array of cross-sections. To plot the contour plots as in Fig.~\ref{fig:contour}, the method \verb|plot_contour()| should be called.

\subsection{Extended mass functions}
\noindent The method \verb|macro.psi(m,param)| should be modified to specify the extended mass function. Similar to the contour constraints, it takes in an array of masses and a list of parameters, returning the value of the distribution $\psi$ as defined in this paper. The mass function should be already normalized, or else it must be imposed by hand (as is shown in the Press-Schechter example in the code comments).

The most important function for general use is \verb|fpsidm(sigma_params,EMF_params)|, which takes in as arguments the parameters for both the cross-section function and the extended mass function. The output of this function is $f_{\psi,\rm DM}$, the fractional constraint on the extended mass function in question. 

If one has $N$ independent distributions to consider, as we demonstrate in the case of our bimodal distribution, one should use \verb|fpsidm_N(sigma_params_list,EMF_params_list)| instead. In this case, each of the arguments is rather a list containing $N$ elements, each of which is the relevant parameter list (e.g. \verb|sigma_params|) for the individual distributions.

The example plots of constraints for extended mass functions that we showed in Figs.~\ref{fig:lognormal},~\ref{fig:ps}, and~\ref{fig:bimodal} were generated with the method \verb|macro.plot_EMF(loop=True)|. This will need suitably modified for the extended mass function in question; the argument \verb|loop=True| determines whether the relevant parameter space of the extended mass function should be looped through to generate $f_{\psi,\mathrm{DM}}$. For convenience the output is saved to a NumPy file after the loop so that one can set \verb|loop=False| after generating the output once and not have to run the time-consuming generation process each time they wish to modify the plots.

\bibliographystyle{bibi}

\bibliography{main.bib}

@article{numpy2020Natur.585..357H,
	title        = {{Array programming with NumPy}},
	author       = {{Harris}, Charles R. and {Millman}, K. Jarrod and {van der Walt}, St{\'e}fan J. and {Gommers}, Ralf and {Virtanen}, Pauli and {Cournapeau}, David and {Wieser}, Eric and {Taylor}, Julian and {Berg}, Sebastian and {Smith}, Nathaniel J. and {Kern}, Robert and {Picus}, Matti and {Hoyer}, Stephan and {van Kerkwijk}, Marten H. and {Brett}, Matthew and {Haldane}, Allan and {del R{\'\i}o}, Jaime Fern{\'a}ndez and {Wiebe}, Mark and {Peterson}, Pearu and {G{\'e}rard-Marchant}, Pierre and {Sheppard}, Kevin and {Reddy}, Tyler and {Weckesser}, Warren and {Abbasi}, Hameer and {Gohlke}, Christoph and {Oliphant}, Travis E.},
	year         = 2020,
	month        = sep,
	journal      = {\nat},
	volume       = 585,
	number       = 7825,
	pages        = {357--362},
	doi          = {10.1038/s41586-020-2649-2},
	archiveprefix = {arXiv},
	eprint       = {2006.10256},
	primaryclass = {cs.MS},
	adsurl       = {https://ui.adsabs.harvard.edu/abs/2020Natur.585..357H}
}

@article{XENON:2023iku,
    author = "Aprile, E. and others",
    collaboration = "XENON",
    title = "{Searching for Heavy Dark Matter near the Planck Mass with XENON1T}",
    eprint = "2304.10931",
    archivePrefix = "arXiv",
    primaryClass = "hep-ex",
    doi = "10.1103/PhysRevLett.130.261002",
    journal = "Phys. Rev. Lett.",
    volume = "130",
    number = "26",
    pages = "261002",
    year = "2023"
}

@article{Acevedo:2021kly,
    author = "Acevedo, Javier F. and Bramante, Joseph and Goodman, Alan",
    title = "{Accelerating composite dark matter discovery with nuclear recoils and the Migdal effect}",
    eprint = "2108.10889",
    archivePrefix = "arXiv",
    primaryClass = "hep-ph",
    doi = "10.1103/PhysRevD.105.023012",
    journal = "Phys. Rev. D",
    volume = "105",
    number = "2",
    pages = "023012",
    year = "2022"
}

@article{Boukhtouchen:2026rfz,
    author = "Boukhtouchen, Yilda and Bramante, Joseph and Buchanan, Andrew and Hayes, Alexander and Leybourne, Matthew and McIntosh, Jennika and Ray, Anupam and Shugar, Aaron",
    title = "{New Windows on Heavy Dark Matter: Mineral Melt Modelling and X-Ray Readout for Muscovite Mica}",
    eprint = "2606.02579",
    archivePrefix = "arXiv",
    primaryClass = "hep-ph",
    month = "6",
    year = "2026"
}

@article{Starkman:1990nj,
    author = "Starkman, Glenn D. and Gould, Andrew and Esmailzadeh, Rahim and Dimopoulos, Savas",
    title = "{Opening the Window on Strongly Interacting Dark Matter}",
    reportNumber = "IASSNS-AST-90/1",
    doi = "10.1103/PhysRevD.41.3594",
    journal = "Phys. Rev. D",
    volume = "41",
    pages = "3594",
    year = "1990"
}

@article{Acevedo:2023cab,
    author = "Acevedo, Javier F. and An, Haipeng and Boukhtouchen, Yilda and Bramante, Joseph and Richardson, Mark L. A. and Sansom, Lucy",
    title = "{Dark matter induced baryonic feedback in galaxies}",
    eprint = "2309.08661",
    archivePrefix = "arXiv",
    primaryClass = "hep-ph",
    doi = "10.1103/PhysRevD.110.083004",
    journal = "Phys. Rev. D",
    volume = "110",
    number = "8",
    pages = "083004",
    year = "2024"
}

@article{Bernabei:1999ui,
    author = "Bernabei, R. and others",
    title = "{Extended limits on neutral strongly interacting massive particles and nuclearites from NaI(Tl) scintillators}",
    doi = "10.1103/PhysRevLett.83.4918",
    journal = "Phys. Rev. Lett.",
    volume = "83",
    pages = "4918--4921",
    year = "1999"
}

@article{LZ:2024psa,
    author = "Aalbers, J. and others",
    collaboration = "LZ",
    title = "{New constraints on ultraheavy dark matter from the LZ experiment}",
    eprint = "2402.08865",
    archivePrefix = "arXiv",
    primaryClass = "hep-ex",
    reportNumber = "FERMILAB-PUB-24-0015-TD",
    doi = "10.1103/PhysRevD.109.112010",
    journal = "Phys. Rev. D",
    volume = "109",
    number = "11",
    pages = "112010",
    year = "2024"
}

@article{mpl4160265,
	title        = {Matplotlib: A 2D Graphics Environment},
	author       = {Hunter, John D.},
	year         = 2007,
	journal      = {Computing in Science and Engineering},
	volume       = 9,
	number       = 3,
	pages        = {90--95},
	doi          = {10.1109/MCSE.2007.55}
}

@article{scipy2020NatMe..17..261V,
	title        = {{SciPy 1.0: fundamental algorithms for scientific computing in Python}},
	author       = {{Virtanen}, Pauli and {Gommers}, Ralf and {Oliphant}, Travis E. and {Haberland}, Matt and {Reddy}, Tyler and {Cournapeau}, David and {Burovski}, Evgeni and {Peterson}, Pearu and {Weckesser}, Warren and {Bright}, Jonathan and {van der Walt}, St{\'e}fan J. and {Brett}, Matthew and {Wilson}, Joshua and {Millman}, K. Jarrod and {Mayorov}, Nikolay and {Nelson}, Andrew R.~J. and {Jones}, Eric and {Kern}, Robert and {Larson}, Eric and {Carey}, C.~J. and {Polat}, {\.I}lhan and {Feng}, Yu and {Moore}, Eric W. and {VanderPlas}, Jake and {Laxalde}, Denis and {Perktold}, Josef and {Cimrman}, Robert and {Henriksen}, Ian and {Quintero}, E.~A. and {Harris}, Charles R. and {Archibald}, Anne M. and {Ribeiro}, Ant{\^o}nio H. and {Pedregosa}, Fabian and {van Mulbregt}, Paul and {SciPy 1. 0 Contributors}},
	year         = 2020,
	month        = feb,
	journal      = {Nature Methods},
	volume       = 17,
	pages        = {261--272},
	doi          = {10.1038/s41592-019-0686-2},
	archiveprefix = {arXiv},
	eprint       = {1907.10121},
	primaryclass = {cs.MS},
	adsurl       = {https://ui.adsabs.harvard.edu/abs/2020NatMe..17..261V}
}

@misc{picker_dark_2025,
    title = {A dark matter hail: {Detecting} macroscopic dark matter with asteroids, planetary rings, and craters},
    shorttitle = {A dark matter hail},
    url = {http://arxiv.org/abs/2504.07232},
    doi = {10.48550/arXiv.2504.07232},
    language = {en},
    urldate = {2025-08-01},
    publisher = {arXiv},
    author = {Picker, Zachary S. C.},
    month = apr,
    year = {2025},
    note = {arXiv:2504.07232 [astro-ph]},
}

@article{gresham_early_2018,
    title = {Early {Universe} synthesis of asymmetric dark matter nuggets},
    volume = {97},
    issn = {2470-0010, 2470-0029},
    url = {https://link.aps.org/doi/10.1103/PhysRevD.97.036003},
    doi = {10.1103/PhysRevD.97.036003},
    language = {en},
    number = {3},
    urldate = {2025-11-14},
    journal = {Physical Review D},
    author = {Gresham, Moira I. and Lou, Hou Keong and Zurek, Kathryn M.},
    month = feb,
    year = {2018},
    pages = {036003},
}

@article{Planck:2018vyg,
    author = "Aghanim, N. and others",
    collaboration = "Planck",
    title = "{Planck 2018 results. VI. Cosmological parameters}",
    eprint = "1807.06209",
    archivePrefix = "arXiv",
    primaryClass = "astro-ph.CO",
    doi = "10.1051/0004-6361/201833910",
    journal = "Astron. Astrophys.",
    volume = "641",
    pages = "A6",
    year = "2020",
    note = "[Erratum: Astron.Astrophys. 652, C4 (2021)]"
}

@article{lu_black_2025,
    title = {Black {Holes} from {Fermi} {Ball} {Collapse}},
    volume = {111},
    issn = {2470-0010, 2470-0029},
    url = {http://arxiv.org/abs/2411.17074},
    doi = {10.1103/PhysRevD.111.043005},
    language = {en},
    number = {4},
    urldate = {2025-11-14},
    journal = {Physical Review D},
    author = {Lu, Yifan and Picker, Zachary S. C. and Profumo, Stefano and Kusenko, Alexander},
    month = feb,
    year = {2025},
    note = {arXiv:2411.17074 [astro-ph]},
    pages = {043005},
}

@article{gresham_nuclear_2017,
    title = {Nuclear structure of bound states of asymmetric dark matter},
    volume = {96},
    copyright = {https://link.aps.org/licenses/aps-default-license},
    issn = {2470-0010, 2470-0029},
    url = {https://link.aps.org/doi/10.1103/PhysRevD.96.096012},
    doi = {10.1103/PhysRevD.96.096012},
    language = {en},
    number = {9},
    urldate = {2025-11-14},
    journal = {Physical Review D},
    author = {Gresham, Moira I. and Lou, Hou Keong and Zurek, Kathryn M.},
    month = nov,
    year = {2017},
    pages = {096012},
}

@article{flores_primordial_2021,
    title = {Primordial black holes from long-range scalar forces and scalar radiative cooling},
    volume = {126},
    issn = {0031-9007, 1079-7114},
    url = {http://arxiv.org/abs/2008.12456},
    doi = {10.1103/PhysRevLett.126.041101},
    language = {en},
    number = {4},
    urldate = {2023-01-19},
    journal = {Physical Review Letters},
    author = {Flores, Marcos M. and Kusenko, Alexander},
    month = jan,
    year = {2021},
    note = {arXiv:2008.12456 [astro-ph, physics:hep-ph]},
    pages = {041101},
}

@misc{flores_gravitational_2022,
    title = {Gravitational waves from rapid structure formation on microscopic scales before matter-radiation equality},
    url = {http://arxiv.org/abs/2209.04970},
    language = {en},
    urldate = {2022-12-05},
    publisher = {arXiv},
    author = {Flores, Marcos M. and Kusenko, Alexander and Sasaki, Misao},
    month = sep,
    year = {2022},
    note = {arXiv:2209.04970 [astro-ph, physics:gr-qc, physics:hep-ph]},
}

@misc{flores_inhomogeneous_2022,
    title = {Inhomogeneous cold electroweak baryogenesis from early structure formation due to {Yukawa} forces},
    url = {http://arxiv.org/abs/2208.09789},
    language = {en},
    urldate = {2022-12-05},
    publisher = {arXiv},
    author = {Flores, Marcos M. and Kusenko, Alexander and Pearce, Lauren and White, Graham},
    month = aug,
    year = {2022},
    note = {arXiv:2208.09789 [astro-ph, physics:hep-ph]},
}

@misc{flores_structure_2023,
    title = {Structure {Formation} after {Reheating}: {Supermassive} {Primordial} {Black} {Holes} and {Fermi} {Ball} {Dark} {Matter}},
    shorttitle = {Structure {Formation} after {Reheating}},
    url = {http://arxiv.org/abs/2308.09094},
    language = {en},
    urldate = {2023-09-06},
    publisher = {arXiv},
    author = {Flores, Marcos M. and Lu, Yifan and Kusenko, Alexander},
    month = aug,
    year = {2023},
    note = {arXiv:2308.09094 [astro-ph, physics:hep-ph]},
}

@misc{domenech_halo_2023,
    title = {Halo {Formation} from {Yukawa} {Forces} in the {Very} {Early} {Universe}},
    url = {http://arxiv.org/abs/2304.13053},
    language = {en},
    urldate = {2023-04-27},
    publisher = {arXiv},
    author = {Domènech, Guillem and Inman, Derek and Kusenko, Alexander and Sasaki, Misao},
    month = apr,
    year = {2023},
    note = {arXiv:2304.13053 [astro-ph, physics:gr-qc, physics:hep-ph, physics:hep-th]},
}

@article{domenech_cosmology_2021,
    title = {Cosmology of strongly interacting fermions in the early universe},
    volume = {2021},
    issn = {1475-7516},
    url = {http://arxiv.org/abs/2104.05271},
    doi = {10.1088/1475-7516/2021/06/030},
    language = {en},
    number = {06},
    urldate = {2026-04-27},
    journal = {Journal of Cosmology and Astroparticle Physics},
    author = {Domènech, Guillem and Sasaki, Misao},
    month = jun,
    year = {2021},
    note = {arXiv:2104.05271 [hep-th]},
    pages = {030},
}

@article{amendola_primordial_2018,
    title = {Primordial black holes from fifth forces},
    volume = {97},
    issn = {2470-0010, 2470-0029},
    url = {https://link.aps.org/doi/10.1103/PhysRevD.97.081302},
    doi = {10.1103/PhysRevD.97.081302},
    language = {en},
    number = {8},
    urldate = {2026-05-01},
    journal = {Physical Review D},
    author = {Amendola, Luca and Rubio, Javier and Wetterich, Christof},
    month = apr,
    year = {2018},
    pages = {081302},
}

@article{savastano_primordial_2019,
    title = {Primordial dark matter halos from fifth forces},
    volume = {100},
    issn = {2470-0010, 2470-0029},
    url = {https://link.aps.org/doi/10.1103/PhysRevD.100.083518},
    doi = {10.1103/PhysRevD.100.083518},
    language = {en},
    number = {8},
    urldate = {2026-05-01},
    journal = {Physical Review D},
    author = {Savastano, Stefano and Amendola, Luca and Rubio, Javier and Wetterich, Christof},
    month = oct,
    year = {2019},
    pages = {083518},
}

@article{Gradwohl:1992ue,
    author = "Gradwohl, Ben-Ami and Frieman, Joshua A.",
    title = "{Dark matter, long range forces, and large scale structure}",
    reportNumber = "FERMILAB-PUB-92-008-A",
    doi = "10.1086/171865",
    journal = "Astrophys. J.",
    volume = "398",
    pages = "407--424",
    year = "1992"
}

@article{nusser_structure_2005,
    title = {Structure formation with a long-range scalar dark matter interaction},
    volume = {71},
    copyright = {http://link.aps.org/licenses/aps-default-license},
    issn = {1550-7998, 1550-2368},
    url = {https://link.aps.org/doi/10.1103/PhysRevD.71.083505},
    doi = {10.1103/PhysRevD.71.083505},
    language = {en},
    number = {8},
    urldate = {2026-05-01},
    journal = {Physical Review D},
    author = {Nusser, Adi and Gubser, S. S. and Peebles, P. J. E.},
    month = apr,
    year = {2005},
    pages = {083505},
}

@article{gubser_structure_2004,
    title = {Structure formation in a string-inspired modification of the cold dark matter model},
    volume = {70},
    copyright = {http://link.aps.org/licenses/aps-default-license},
    issn = {1550-7998, 1550-2368},
    url = {https://link.aps.org/doi/10.1103/PhysRevD.70.123510},
    doi = {10.1103/PhysRevD.70.123510},
    language = {en},
    number = {12},
    urldate = {2026-05-01},
    journal = {Physical Review D},
    author = {Gubser, Steven S. and Peebles, P. J. E.},
    month = dec,
    year = {2004},
    pages = {123510},
}

@misc{macrolimits,
  author       = {Zachary S. C. Picker},
  title        = {zpicker/macrolimits: Macroscopic Dark Matter Limits},
  month        = aug,
  year         = 2026,
  publisher    = {Zenodo},
  version      = {v1.0},
  doi          = {10.5281/zenodo.21892125},
  howpublished = {\url{https://github.com/zpicker/macrolimits}}
}

@misc{oharedd,
  author       = {Ciaran A.J. O'Hare},
  title        = {cajohare/DirectDetectionPlots},
  month        = oct,
  year         = 2024,
  publisher    = {Zenodo},
  version      = {v1.0},
  howpublished = {\url{https://github.com/cajohare/DirectDetectionPlots}}
}

@misc{kavanaghpbh,
  author       = {Bradley J. Kavanagh},
  title        = {bradkav/PBHbounds},
  month        = Nov,
  year         = 2019,
  publisher    = {Zenodo},
  version      = {v1.0},
  howpublished = {\url{https://github.com/bradkav/PBHbounds}}
  }

@article{rafelski_compact_2012,
    title = {Compact {Ultradense} {Objects} in the {Solar} {System}},
    volume = {43},
    issn = {0587-4254, 1509-5770},
    url = {http://arxiv.org/abs/1303.4506},
    doi = {10.5506/APhysPolB.43.2251},
    language = {en},
    number = {12},
    urldate = {2026-08-11},
    journal = {Acta Physica Polonica B},
    author = {Rafelski, J. and Dietl, Ch and Labun, L.},
    year = {2012},
    note = {arXiv:1303.4506 [astro-ph.EP]},
    pages = {2251},
}

@article{rafelski_compact_2013,
    title = {Compact {Ultra} {Dense} {Matter} {Impactors}},
    volume = {110},
    issn = {0031-9007, 1079-7114},
    url = {http://arxiv.org/abs/1104.4572},
    doi = {10.1103/PhysRevLett.110.111102},
    language = {en},
    number = {11},
    urldate = {2026-08-11},
    journal = {Physical Review Letters},
    author = {Rafelski, Johann and Labun, Lance and Birrell, Jeremiah},
    month = mar,
    year = {2013},
    note = {arXiv:1104.4572 [astro-ph.EP]},
    pages = {111102},
}

@article{griest_galactic_1991,
    title = {Galactic microlensing as a method of detecting massive compact halo objects},
    volume = {366},
    issn = {0004-637X, 1538-4357},
    url = {http://adsabs.harvard.edu/doi/10.1086/169575},
    doi = {10.1086/169575},
    language = {en},
    urldate = {2026-08-11},
    journal = {The Astrophysical Journal},
    author = {Griest, Kim},
    month = jan,
    year = {1991},
    pages = {412},
}

@article{witten_cosmic_1984,
    title = {Cosmic separation of phases},
    volume = {30},
    issn = {0556-2821},
    url = {https://link.aps.org/doi/10.1103/PhysRevD.30.272},
    doi = {10.1103/PhysRevD.30.272},
    language = {en},
    number = {2},
    urldate = {2022-12-07},
    journal = {Physical Review D},
    author = {Witten, Edward},
    month = jul,
    year = {1984},
    pages = {272--285},
}

@article{de_rujula_nuclearitesnovel_1984,
    title = {Nuclearites—a novel form of cosmic radiation},
    volume = {312},
    issn = {0028-0836, 1476-4687},
    url = {http://www.nature.com/articles/312734a0},
    doi = {10.1038/312734a0},
    language = {en},
    number = {5996},
    urldate = {2023-01-10},
    journal = {Nature},
    author = {De Rújula, A. and Glashow, S. L.},
    month = dec,
    year = {1984},
    pages = {734--737},
}

@article{farhi_strange_1984,
    title = {Strange matter},
    volume = {30},
    issn = {0556-2821},
    url = {https://link.aps.org/doi/10.1103/PhysRevD.30.2379},
    doi = {10.1103/PhysRevD.30.2379},
    language = {en},
    number = {11},
    urldate = {2022-12-16},
    journal = {Physical Review D},
    author = {Farhi, Edward and Jaffe, R. L.},
    month = dec,
    year = {1984},
    pages = {2379--2390},
}

@article{alcock_evaporation_1985,
    title = {Evaporation of strange matter in the early {Universe}},
    volume = {32},
    issn = {0556-2821},
    url = {https://link.aps.org/doi/10.1103/PhysRevD.32.1273},
    doi = {10.1103/PhysRevD.32.1273},
    language = {en},
    number = {6},
    urldate = {2022-12-16},
    journal = {Physical Review D},
    author = {Alcock, Charles and Farhi, Edward},
    month = sep,
    year = {1985},
    pages = {1273--1279},
}

@misc{kolb_wimpzillas_1998,
    title = {{WIMPZILLAS}!},
    url = {http://arxiv.org/abs/hep-ph/9810361},
    doi = {10.48550/arXiv.hep-ph/9810361},
    language = {en},
    urldate = {2026-08-11},
    publisher = {arXiv},
    author = {Kolb, Edward W. and Chung, Daniel J. H. and Riotto, Antonio},
    month = oct,
    year = {1998},
    note = {arXiv:hep-ph/9810361},
}

@article{chung_nonthermal_1998,
    title = {Nonthermal {Supermassive} {Dark} {Matter}},
    volume = {81},
    issn = {0031-9007, 1079-7114},
    url = {http://arxiv.org/abs/hep-ph/9805473},
    doi = {10.1103/PhysRevLett.81.4048},
    language = {en},
    number = {19},
    urldate = {2026-08-11},
    journal = {Physical Review Letters},
    author = {Chung, Daniel J. H. and Kolb, Edward W. and Riotto, Antonio},
    month = nov,
    year = {1998},
    note = {arXiv:hep-ph/9805473},
    pages = {4048--4051},
}

@article{kuzmin_ultra-high_1998,
    title = {Ultra-{High} {Energy} {Cosmic} {Rays}, {Superheavy} {Long}-{Living} {Particles}, and {Matter} {Creation} after {Inflation}},
    volume = {68},
    issn = {0021-3640, 1090-6487},
    url = {http://arxiv.org/abs/hep-ph/9802304},
    doi = {10.1134/1.567858},
    language = {en},
    number = {4},
    urldate = {2026-08-11},
    journal = {Journal of Experimental and Theoretical Physics Letters},
    author = {Kuzmin, V. A. and Tkachev, I. I.},
    month = aug,
    year = {1998},
    note = {arXiv:hep-ph/9802304},
    pages = {271--275},
}

@article{dhakal_new_2023,
    title = {New constraints on macroscopic dark matter using radar meteor detectors},
    volume = {107},
    issn = {2470-0010, 2470-0029},
    url = {https://link.aps.org/doi/10.1103/PhysRevD.107.043026},
    doi = {10.1103/PhysRevD.107.043026},
    language = {en},
    number = {4},
    urldate = {2026-08-11},
    journal = {Physical Review D},
    author = {Dhakal, Pawan and Prohira, Steven and Cappiello, Christopher V. and Beacom, John F. and Palo, Scott and Marino, John},
    month = feb,
    year = {2023},
    pages = {043026},
}

@article{starkman_straight_2021,
    title = {Straight {Lightning} as a {Signature} of {Macroscopic} {Dark} {Matter}},
    volume = {103},
    issn = {2470-0010, 2470-0029},
    url = {http://arxiv.org/abs/2006.16272},
    doi = {10.1103/PhysRevD.103.063024},
    language = {en},
    number = {6},
    urldate = {2026-08-11},
    journal = {Physical Review D},
    author = {Starkman, Nathaniel and Sidhu, Jagjit and Winch, Harrison and Starkman, Glenn},
    month = mar,
    year = {2021},
    note = {arXiv:2006.16272 [astro-ph.CO]},
    pages = {063024},
}

@article{sidhu_death_2020,
    title = {Death and {Serious} {Injury} by {Dark} {Matter}},
    volume = {803},
    issn = {03702693},
    url = {http://arxiv.org/abs/1907.06674},
    doi = {10.1016/j.physletb.2020.135300},
    language = {en},
    urldate = {2026-08-11},
    journal = {Physics Letters B},
    author = {Sidhu, Jagjit Singh and Scherrer, Robert J. and Starkman, Glenn},
    month = apr,
    year = {2020},
    note = {arXiv:1907.06674 [astro-ph.CO]},
    pages = {135300},
}

@article{Bhoonah:2018wmw,
    author = "Bhoonah, Amit and Bramante, Joseph and Elahi, Fatemeh and Schon, Sarah",
    title = "{Calorimetric Dark Matter Detection With Galactic Center Gas Clouds}",
    eprint = "1806.06857",
    archivePrefix = "arXiv",
    primaryClass = "hep-ph",
    doi = "10.1103/PhysRevLett.121.131101",
    journal = "Phys. Rev. Lett.",
    volume = "121",
    number = "13",
    pages = "131101",
    year = "2018"
}

@article{yuan2009annealing,
  title={Annealing behavior of alpha recoil tracks in phlogopite},
  author={Yuan, Wanming and Ketcham, Richard A and Gao, Shaokai and Dong, Jinquan and Bao, Zenkuan and Deng, Jun},
  journal={Chemical Geology},
  volume={266},
  number={3-4},
  pages={343--349},
  year={2009},
  publisher={Elsevier}
}

@article{Bramante:2024hbr,
    author = "Bramante, Joseph and Diamond, Melissa D. and Kim, J. Leo",
    title = "{Dimming Starlight with Dark Compact Objects}",
    eprint = "2409.08322",
    archivePrefix = "arXiv",
    primaryClass = "hep-ph",
    doi = "10.1103/PhysRevLett.134.141001",
    journal = "Phys. Rev. Lett.",
    volume = "134",
    number = "14",
    pages = "141001",
    year = "2025"
}

@article{Bhoonah:2018gjb,
    author = "Bhoonah, Amit and Bramante, Joseph and Elahi, Fatemeh and Schon, Sarah",
    title = "{Galactic Center gas clouds and novel bounds on ultralight dark photon, vector portal, strongly interacting, composite, and super-heavy dark matter}",
    eprint = "1812.10919",
    archivePrefix = "arXiv",
    primaryClass = "hep-ph",
    doi = "10.1103/PhysRevD.100.023001",
    journal = "Phys. Rev. D",
    volume = "100",
    number = "2",
    pages = "023001",
    year = "2019"
}

@article{bhoonah_detecting_2021,
    title = {Detecting {Composite} {Dark} {Matter} with {Long} {Range} and {Contact} {Interactions} in {Gas} {Clouds}},
    volume = {103},
    issn = {2470-0010, 2470-0029},
    url = {http://arxiv.org/abs/2010.07240},
    doi = {10.1103/PhysRevD.103.123026},
    language = {en},
    number = {12},
    urldate = {2026-08-11},
    journal = {Physical Review D},
    author = {Bhoonah, Amit and Bramante, Joseph and Schon, Sarah and Song, Ningqiang},
    month = jun,
    year = {2021},
    note = {arXiv:2010.07240 [hep-ph]},
    pages = {123026},
}

@article{bagnasco_detecting_1994,
    title = {Detecting {Technibaryon} {Dark} {Matter}},
    volume = {320},
    issn = {03702693},
    url = {http://arxiv.org/abs/hep-ph/9310290},
    doi = {10.1016/0370-2693(94)90830-3},
    language = {en},
    number = {1-2},
    urldate = {2026-08-11},
    journal = {Physics Letters B},
    author = {Bagnasco, John and Dine, Michael and Thomas, Scott},
    month = jan,
    year = {1994},
    note = {arXiv:hep-ph/9310290},
    pages = {99--104},
}

@article{foadi_technicolor_2009,
    title = {Technicolor {Dark} {Matter}},
    volume = {80},
    issn = {1550-7998, 1550-2368},
    url = {http://arxiv.org/abs/0812.3406},
    doi = {10.1103/PhysRevD.80.037702},
    language = {en},
    number = {3},
    urldate = {2026-08-11},
    journal = {Physical Review D},
    author = {Foadi, Roshan and Frandsen, Mads T. and Sannino, Francesco},
    month = aug,
    year = {2009},
    note = {arXiv:0812.3406 [hep-ph]},
    pages = {037702},
}

@article{kribs_quirky_2010,
    title = {Quirky {Composite} {Dark} {Matter}},
    volume = {81},
    issn = {1550-7998, 1550-2368},
    url = {http://arxiv.org/abs/0909.2034},
    doi = {10.1103/PhysRevD.81.095001},
    language = {en},
    number = {9},
    urldate = {2026-08-11},
    journal = {Physical Review D},
    author = {Kribs, Graham D. and Roy, Tuhin S. and Terning, John and Zurek, Kathryn M.},
    month = may,
    year = {2010},
    note = {arXiv:0909.2034 [hep-ph]},
    pages = {095001},
}

@article{detmold_dark_2014,
    title = {Dark {Nuclei} {I}: {Cosmology} and {Indirect} {Detection}},
    volume = {90},
    issn = {1550-7998, 1550-2368},
    shorttitle = {Dark {Nuclei} {I}},
    url = {http://arxiv.org/abs/1406.2276},
    doi = {10.1103/PhysRevD.90.115013},
    language = {en},
    number = {11},
    urldate = {2026-08-11},
    journal = {Physical Review D},
    author = {Detmold, William and McCullough, Matthew and Pochinsky, Andrew},
    month = dec,
    year = {2014},
    note = {arXiv:1406.2276 [hep-ph]},
    pages = {115013},
}

@article{krnjaic_big_2015,
    title = {Big {Bang} {Darkleosynthesis}},
    volume = {751},
    issn = {03702693},
    url = {http://arxiv.org/abs/1406.1171},
    doi = {10.1016/j.physletb.2015.11.001},
    language = {en},
    urldate = {2026-08-11},
    journal = {Physics Letters B},
    author = {Krnjaic, Gordan and Sigurdson, Kris},
    month = dec,
    year = {2015},
    note = {arXiv:1406.1171 [hep-ph]},
    pages = {464--468},
}

@article{zhitnitsky_nonbaryonic_2003,
    title = {"{Nonbaryonic}" {Dark} {Matter} as {Baryonic} {Color} {Superconductor}},
    volume = {2003},
    issn = {1475-7516},
    url = {http://arxiv.org/abs/hep-ph/0202161},
    doi = {10.1088/1475-7516/2003/10/010},
    language = {en},
    number = {10},
    urldate = {2026-08-11},
    journal = {Journal of Cosmology and Astroparticle Physics},
    author = {Zhitnitsky, Ariel R.},
    month = oct,
    year = {2003},
    note = {arXiv:hep-ph/0202161},
    pages = {010--010},
}

@article{wise_stable_2014,
    title = {Stable {Bound} {States} of {Asymmetric} {Dark} {Matter}},
    volume = {90},
    issn = {1550-7998, 1550-2368},
    url = {http://arxiv.org/abs/1407.4121},
    doi = {10.1103/PhysRevD.90.055030},
    language = {en},
    number = {5},
    urldate = {2026-08-11},
    journal = {Physical Review D},
    author = {Wise, Mark B. and Zhang, Yue},
    month = sep,
    year = {2014},
    note = {arXiv:1407.4121 [hep-ph]},
    pages = {055030},
}

@article{wise_yukawa_2015,
    title = {Yukawa {Bound} {States} of a {Large} {Number} of {Fermions}},
    volume = {2015},
    issn = {1029-8479},
    url = {http://arxiv.org/abs/1411.1772},
    doi = {10.1007/JHEP02(2015)023},
    language = {en},
    number = {2},
    urldate = {2026-08-11},
    journal = {Journal of High Energy Physics},
    author = {Wise, Mark B. and Zhang, Yue},
    month = feb,
    year = {2015},
    note = {arXiv:1411.1772 [hep-ph]},
    pages = {23},
}

@article{hardy_big_2015,
    title = {Big {Bang} {Synthesis} of {Nuclear} {Dark} {Matter}},
    volume = {2015},
    issn = {1029-8479},
    url = {http://arxiv.org/abs/1411.3739},
    doi = {10.1007/JHEP06(2015)011},
    language = {en},
    number = {6},
    urldate = {2026-08-11},
    journal = {Journal of High Energy Physics},
    author = {Hardy, Edward and Lasenby, Robert and March-Russell, John and West, Stephen M.},
    month = jun,
    year = {2015},
    note = {arXiv:1411.3739 [hep-ph]},
    pages = {11},
}

@article{jacobs_macro_2015,
    title = {Macro {Dark} {Matter}},
    volume = {450},
    issn = {1365-2966, 0035-8711},
    url = {http://arxiv.org/abs/1410.2236},
    doi = {10.1093/mnras/stv774},
    language = {en},
    number = {4},
    urldate = {2026-08-11},
    journal = {Monthly Notices of the Royal Astronomical Society},
    author = {Jacobs, David M. and Starkman, Glenn D. and Lynn, Bryan W.},
    month = jul,
    year = {2015},
    note = {arXiv:1410.2236 [astro-ph.CO]},
    pages = {3418--3430},
}

@article{ge_cosmological_2018,
    title = {Cosmological {Axion} and {Quark} {Nugget} {Dark} {Matter} {Model}},
    volume = {97},
    issn = {2470-0010, 2470-0029},
    url = {http://arxiv.org/abs/1711.06271},
    doi = {10.1103/PhysRevD.97.043008},
    language = {en},
    number = {4},
    urldate = {2026-08-11},
    journal = {Physical Review D},
    author = {Ge, Shuailiang and Liang, Xunyu and Zhitnitsky, Ariel},
    month = feb,
    year = {2018},
    note = {arXiv:1711.06271 [hep-ph]},
    pages = {043008},
}

@article{bai_dark_2019,
    title = {Dark {Quark} {Nuggets}},
    volume = {99},
    issn = {2470-0010, 2470-0029},
    url = {http://arxiv.org/abs/1810.04360},
    doi = {10.1103/PhysRevD.99.055047},
    language = {en},
    number = {5},
    urldate = {2026-08-11},
    journal = {Physical Review D},
    author = {Bai, Yang and Long, Andrew J. and Lu, Sida},
    month = mar,
    year = {2019},
    note = {arXiv:1810.04360 [hep-ph]},
    pages = {055047},
}

@article{bramante_saturated_2018,
    title = {Saturated {Overburden} {Scattering} and the {Multiscatter} {Frontier}: {Discovering} {Dark} {Matter} at the {Planck} {Mass} and {Beyond}},
    volume = {98},
    issn = {2470-0010, 2470-0029},
    shorttitle = {Saturated {Overburden} {Scattering} and the {Multiscatter} {Frontier}},
    url = {http://arxiv.org/abs/1803.08044},
    doi = {10.1103/PhysRevD.98.083516},
    language = {en},
    number = {8},
    urldate = {2026-08-11},
    journal = {Physical Review D},
    author = {Bramante, Joseph and Broerman, Benjamin and Lang, Rafael F. and Raj, Nirmal},
    month = oct,
    year = {2018},
    note = {arXiv:1803.08044 [hep-ph]},
    pages = {083516},
}

@article{Espriu:2026jzi,
    author = "Espriu, Domenec",
    title = "{Do primordial quark pellets solve the dark matter puzzle?}",
    eprint = "2607.10672",
    archivePrefix = "arXiv",
    primaryClass = "hep-ph",
    month = "7",
    year = "2026"
}

@misc{acevedo_loosely_2024,
    title = {Loosely {Bound} {Composite} {Dark} {Matter}},
    url = {http://arxiv.org/abs/2408.03983},
    doi = {10.48550/arXiv.2408.03983},
    language = {en},
    urldate = {2025-11-24},
    publisher = {arXiv},
    author = {Acevedo, Javier F. and Boukhtouchen, Yilda and Bramante, Joseph and Cappiello, Chris and Mohlabeng, Gopolang and Tyagi, Narayani},
    month = aug,
    year = {2024},
    note = {arXiv:2408.03983 [hep-ph]},
}

@misc{bramante_very_2026,
    title = {Very {Heavy} and {Composite} {Dark} {Matter}: {Theory} and {Experimental} {Searches}},
    shorttitle = {Very {Heavy} and {Composite} {Dark} {Matter}},
    url = {http://arxiv.org/abs/2602.23708},
    doi = {10.48550/arXiv.2602.23708},
    language = {en},
    urldate = {2026-08-11},
    publisher = {arXiv},
    author = {Bramante, Joseph},
    month = feb,
    year = {2026},
    note = {arXiv:2602.23708 [hep-ph]},
}

@article{lynn_strange_1990,
    title = {Strange baryon matter},
    volume = {345},
    issn = {05503213},
    url = {https://linkinghub.elsevier.com/retrieve/pii/055032139090614J},
    doi = {10.1016/0550-3213(90)90614-J},
    language = {en},
    number = {1},
    urldate = {2026-08-11},
    journal = {Nuclear Physics B},
    author = {Lynn, Bryan W. and Nelson, Ann E. and Tetradis, Nikolaos},
    month = nov,
    year = {1990},
    pages = {186--209},
}

@article{zhitnitsky_cold_2006,
    title = {Cold dark matter as compact composite objects},
    volume = {74},
    issn = {1550-7998, 1550-2368},
    url = {https://link.aps.org/doi/10.1103/PhysRevD.74.043515},
    doi = {10.1103/PhysRevD.74.043515},
    language = {en},
    number = {4},
    urldate = {2022-12-16},
    journal = {Physical Review D},
    author = {Zhitnitsky, Ariel},
    month = aug,
    year = {2006},
    pages = {043515},
}

@article{lee_fermion_1987,
    title = {Fermion soliton stars and black holes},
    volume = {35},
    copyright = {http://link.aps.org/licenses/aps-default-license},
    issn = {0556-2821},
    url = {https://link.aps.org/doi/10.1103/PhysRevD.35.3678},
    doi = {10.1103/PhysRevD.35.3678},
    language = {en},
    number = {12},
    urldate = {2026-08-11},
    journal = {Physical Review D},
    author = {Lee, T. D. and Pang, Y.},
    month = jun,
    year = {1987},
    pages = {3678--3694},
}

@article{coleman_q-balls_1985,
    title = {Q-balls},
    volume = {262},
    copyright = {https://www.elsevier.com/tdm/userlicense/1.0/},
    issn = {05503213},
    url = {https://linkinghub.elsevier.com/retrieve/pii/055032138590286X},
    doi = {10.1016/0550-3213(85)90286-X},
    language = {en},
    number = {2},
    urldate = {2026-08-11},
    journal = {Nuclear Physics B},
    author = {Coleman, Sidney},
    month = dec,
    year = {1985},
    pages = {263--283},
}

@article{kusenko_supersymmetric_1998,
    title = {Supersymmetric {Q}-balls as dark matter},
    volume = {418},
    copyright = {https://www.elsevier.com/tdm/userlicense/1.0/},
    issn = {03702693},
    url = {https://linkinghub.elsevier.com/retrieve/pii/S0370269397013750},
    doi = {10.1016/S0370-2693(97)01375-0},
    language = {en},
    number = {1-2},
    urldate = {2026-08-11},
    journal = {Physics Letters B},
    author = {Kusenko, Alexander and Shaposhnikov, Mikhail},
    month = jan,
    year = {1998},
    pages = {46--54},
}

@article{macpherson_biased_1995,
    title = {Biased discrete symmetry breaking and {Fermi} balls},
    volume = {347},
    copyright = {https://www.elsevier.com/tdm/userlicense/1.0/},
    issn = {03702693},
    url = {https://linkinghub.elsevier.com/retrieve/pii/0370269395000805},
    doi = {10.1016/0370-2693(95)00080-5},
    language = {en},
    number = {3-4},
    urldate = {2026-08-11},
    journal = {Physics Letters B},
    author = {Macpherson, Alick L and Campbell, Bruce A},
    month = mar,
    year = {1995},
    pages = {205--210},
}

@article{pbh,
	title        = {{The hypothesis of cores retarded during expansion and the hot cosmological model}},
	author       = {Zel'dovich, Ya. B. and Novikov, I. D.},
	year         = 1966,
	journal      = {Sov. Astron.},
	volume       = 10,
	pages        = 602,
	adsurl       = {https://ui.adsabs.harvard.edu/abs/1967SvA....10..602Z}
}

@article{Hawking:1971ei,
	title        = {{Gravitationally collapsed objects of very low mass}},
	author       = {Hawking, Stephen},
	year         = 1971,
	journal      = {Mon. Not. Roy. Astron. Soc.},
	volume       = 152,
	pages        = 75,
	doi          = {10.1093/mnras/152.1.75}
}

@article{Carr:1974nx,
	title        = {{Black holes in the early Universe}},
	author       = {Carr, Bernard J. and Hawking, S.W.},
	year         = 1974,
	journal      = {Mon. Not. Roy. Astron. Soc.},
	volume       = 168,
	pages        = {399--415},
	doi          = {10.1093/mnras/168.2.399}
}

@article{Chapline:1975ojl,
	title        = {{Cosmological effects of primordial black holes}},
	author       = {Chapline, George F.},
	year         = 1975,
	journal      = {Nature},
	volume       = 253,
	number       = 5489,
	pages        = {251--252},
	doi          = {10.1038/253251a0}
}

@article{acevedo_dark_2024,
    title = {Dark {Matter}-{Induced} {Baryonic} {Feedback} in {Galaxies}},
    volume = {110},
    issn = {2470-0010, 2470-0029},
    url = {http://arxiv.org/abs/2309.08661},
    doi = {10.1103/PhysRevD.110.083004},
    language = {en},
    number = {8},
    urldate = {2026-08-12},
    journal = {Physical Review D},
    author = {Acevedo, Javier F. and An, Haipeng and Boukhtouchen, Yilda and Bramante, Joseph and Richardson, Mark and Sansom, Lucy},
    month = oct,
    year = {2024},
    note = {arXiv:2309.08661 [hep-ph]},
    pages = {083004},
}

@misc{boukhtouchen_deconstructive_2026,
    title = {Deconstructive {Composite} {Dark} {Matter} {Detection}},
    url = {http://arxiv.org/abs/2512.16043},
    doi = {10.48550/arXiv.2512.16043},
    language = {en},
    urldate = {2026-08-12},
    publisher = {arXiv},
    author = {Boukhtouchen, Yilda and Bramante, Joseph and Cappiello, Christopher and Diamond, Melissa},
    month = feb,
    year = {2026},
    note = {arXiv:2512.16043 [hep-ph]},
}

@article{acevedo_loosely_2025,
    title = {Loosely bound composite dark matter},
    volume = {2025},
    issn = {1475-7516},
    url = {https://iopscience.iop.org/article/10.1088/1475-7516/2025/03/013},
    doi = {10.1088/1475-7516/2025/03/013},
    language = {en},
    number = {03},
    urldate = {2026-08-12},
    journal = {Journal of Cosmology and Astroparticle Physics},
    author = {Acevedo, Javier F. and Boukhtouchen, Yilda and Bramante, Joseph and Cappiello, Christopher and Mohlabeng, Gopolang and Tyagi, Narayani},
    month = mar,
    year = {2025},
    pages = {013},
}

@article{bleau_secluded_2025,
    title = {Secluded dark composites and remnant binding fields},
    volume = {2025},
    issn = {1475-7516},
    url = {https://iopscience.iop.org/article/10.1088/1475-7516/2025/12/034},
    doi = {10.1088/1475-7516/2025/12/034},
    language = {en},
    number = {12},
    urldate = {2026-08-12},
    journal = {Journal of Cosmology and Astroparticle Physics},
    author = {Bleau, Katarina and Boukhtouchen, Yilda and Bramante, Joseph and Kulkarni, Rohan},
    month = dec,
    year = {2025},
    pages = {034},
}

@article{bottkejr_linking_2005,
    title = {Linking the collisional history of the main asteroid belt to its dynamical excitation and depletion},
    volume = {179},
    copyright = {https://www.elsevier.com/tdm/userlicense/1.0/},
    issn = {00191035},
    url = {https://linkinghub.elsevier.com/retrieve/pii/S0019103505001958},
    doi = {10.1016/j.icarus.2005.05.017},
    language = {en},
    number = {1},
    urldate = {2025-01-29},
    journal = {Icarus},
    author = {Bottkejr, W and Durda, D and Nesvorny, D and Jedicke, R and Morbidelli, A and Vokrouhlicky, D and Levison, H},
    month = dec,
    year = {2005},
    pages = {63--94},
}

@article{crida_age_2025,
    title = {The {Age} and {Origin} of {Saturn}’s {Rings}},
    volume = {221},
    issn = {0038-6308, 1572-9672},
    url = {https://link.springer.com/10.1007/s11214-025-01189-z},
    doi = {10.1007/s11214-025-01189-z},
    language = {en},
    number = {5},
    urldate = {2025-09-23},
    journal = {Space Science Reviews},
    author = {Crida, Aurélien and Estrada, Paul R. and Nicholson, Philip D. and Murray, Carl D.},
    month = aug,
    year = {2025},
    pages = {66},
}

@article{cappiello_new_2021,
    title = {New {Experimental} {Constraints} in a {New} {Landscape} for {Composite} {Dark} {Matter}},
    volume = {103},
    issn = {2470-0010, 2470-0029},
    url = {http://arxiv.org/abs/2008.10646},
    doi = {10.1103/PhysRevD.103.023019},
    language = {en},
    number = {2},
    urldate = {2026-08-17},
    journal = {Physical Review D},
    author = {Cappiello, Christopher V. and Collar, J. I. and Beacom, John F.},
    month = jan,
    year = {2021},
    note = {arXiv:2008.10646 [hep-ex]},
    pages = {023019},
}

@article{Collar:2018ydf,
    author = "Collar, J. I.",
    title = "{Search for a nonrelativistic component in the spectrum of cosmic rays at Earth}",
    eprint = "1805.02646",
    archivePrefix = "arXiv",
    primaryClass = "astro-ph.CO",
    doi = "10.1103/PhysRevD.98.023005",
    journal = "Phys. Rev. D",
    volume = "98",
    number = "2",
    pages = "023005",
    year = "2018"
}

@article{Blanco:2019lrf,
    author = "Blanco, Carlos and Collar, J. I. and Kahn, Yonatan and Lillard, Benjamin",
    title = "{Dark Matter-Electron Scattering from Aromatic Organic Targets}",
    eprint = "1912.02822",
    archivePrefix = "arXiv",
    primaryClass = "hep-ph",
    doi = "10.1103/PhysRevD.101.056001",
    journal = "Phys. Rev. D",
    volume = "101",
    number = "5",
    pages = "056001",
    year = "2020"
}

@article{Clark:2020mna,
    author = "Clark, Michael and Depoian, Amanda and Elshimy, Bahaa and Kopec, Abigail and Lang, Rafael F. and Li, Shengchao and Qin, Juehang",
    title = "{Direct Detection Limits on Heavy Dark Matter}",
    eprint = "2009.07909",
    archivePrefix = "arXiv",
    primaryClass = "hep-ph",
    doi = "10.1103/PhysRevD.102.123026",
    journal = "Phys. Rev. D",
    volume = "102",
    number = "12",
    pages = "123026",
    year = "2020"
}

@article{XENON:2018voc,
    author = "Aprile, E. and others",
    collaboration = "XENON",
    title = "{Dark Matter Search Results from a One Ton-Year Exposure of XENON1T}",
    eprint = "1805.12562",
    archivePrefix = "arXiv",
    primaryClass = "astro-ph.CO",
    doi = "10.1103/PhysRevLett.121.111302",
    journal = "Phys. Rev. Lett.",
    volume = "121",
    number = "11",
    pages = "111302",
    year = "2018"
}

@article{CDMS:2000lgz,
    author = "Abusaidi, R. and others",
    collaboration = "CDMS",
    title = "{Exclusion limits on the WIMP nucleon cross-section from the cryogenic dark matter search}",
    eprint = "astro-ph/0002471",
    archivePrefix = "arXiv",
    reportNumber = "CWRU-P5-00, UCSB-HEP-00-01, FERMILAB-PUB-00-384-AD-E, CWRU-P5-00-UCSB-HEP-00-01",
    doi = "10.1103/PhysRevLett.84.5699",
    journal = "Phys. Rev. Lett.",
    volume = "84",
    pages = "5699--5703",
    year = "2000"
}

@article{CDMS:2002moo,
    author = "Abrams, D. and others",
    collaboration = "CDMS",
    title = "{Exclusion Limits on the WIMP Nucleon Cross-Section from the Cryogenic Dark Matter Search}",
    eprint = "astro-ph/0203500",
    archivePrefix = "arXiv",
    reportNumber = "FERMILAB-PUB-02-287, CWRU-P4-02, UCSB-HEP-02-02, CWRU-P4-02-UCSB-HEP-02-02",
    doi = "10.1103/PhysRevD.66.122003",
    journal = "Phys. Rev. D",
    volume = "66",
    pages = "122003",
    year = "2002"
}

@article{Kavanagh:2017cru,
    author = "Kavanagh, Bradley J.",
    title = "{Earth scattering of superheavy dark matter: Updated constraints from detectors old and new}",
    eprint = "1712.04901",
    archivePrefix = "arXiv",
    primaryClass = "hep-ph",
    doi = "10.1103/PhysRevD.97.123013",
    journal = "Phys. Rev. D",
    volume = "97",
    number = "12",
    pages = "123013",
    year = "2018"
}

@article{DEAPCollaboration:2021raj,
    author = "Adhikari, P. and others",
    collaboration = "(DEAP Collaboration){\textdaggerdbl}, DEAP",
    title = "{First Direct Detection Constraints on Planck-Scale Mass Dark Matter with Multiple-Scatter Signatures Using the DEAP-3600 Detector}",
    eprint = "2108.09405",
    archivePrefix = "arXiv",
    primaryClass = "astro-ph.CO",
    doi = "10.1103/PhysRevLett.128.011801",
    journal = "Phys. Rev. Lett.",
    volume = "128",
    number = "1",
    pages = "011801",
    year = "2022"
}

@article{DEAP-3600:2017ker,
    author = "Amaudruz, P. -A. and others",
    collaboration = "DEAP-3600",
    title = "{Design and Construction of the DEAP-3600 Dark Matter Detector}",
    eprint = "1712.01982",
    archivePrefix = "arXiv",
    primaryClass = "astro-ph.IM",
    doi = "10.1016/j.astropartphys.2018.09.006",
    journal = "Astropart. Phys.",
    volume = "108",
    pages = "1--23",
    year = "2019"
}

@article{Sugiyama:2026kpv,
    author = "Sugiyama, Sunao and Takada, Masahiro and Yasuda, Naoki and Tominaga, Nozomu",
    title = "{Microlensing constraints on Primordial Black Hole abundance with Subaru Hyper Suprime-Cam observations of Andromeda}",
    eprint = "2602.05840",
    archivePrefix = "arXiv",
    primaryClass = "astro-ph.CO",
    month = "2",
    year = "2026"
}

@article{Croon:2020wpr,
    author = "Croon, Djuna and McKeen, David and Raj, Nirmal",
    title = "{Gravitational microlensing by dark matter in extended structures}",
    eprint = "2002.08962",
    archivePrefix = "arXiv",
    primaryClass = "astro-ph.CO",
    doi = "10.1103/PhysRevD.101.083013",
    journal = "Phys. Rev. D",
    volume = "101",
    number = "8",
    pages = "083013",
    year = "2020"
}

@article{Dvorkin:2013cea,
    author = "Dvorkin, Cora and Blum, Kfir and Kamionkowski, Marc",
    title = "{Constraining Dark Matter-Baryon Scattering with Linear Cosmology}",
    eprint = "1311.2937",
    archivePrefix = "arXiv",
    primaryClass = "astro-ph.CO",
    doi = "10.1103/PhysRevD.89.023519",
    journal = "Phys. Rev. D",
    volume = "89",
    number = "2",
    pages = "023519",
    year = "2014"
}

@article{Gluscevic:2017ywp,
    author = "Gluscevic, Vera and Boddy, Kimberly K.",
    title = "{Constraints on Scattering of keV{\textendash}TeV Dark Matter with Protons in the Early Universe}",
    eprint = "1712.07133",
    archivePrefix = "arXiv",
    primaryClass = "astro-ph.CO",
    doi = "10.1103/PhysRevLett.121.081301",
    journal = "Phys. Rev. Lett.",
    volume = "121",
    number = "8",
    pages = "081301",
    year = "2018"
}

@article{Chivukula:1989cc,
    author = "Chivukula, R. Sekhar and Cohen, Andrew G. and Dimopoulos, Savas and Walker, Terry P.",
    title = "{Bounds on Halo Particle Interactions From Interstellar Calorimetry}",
    reportNumber = "BUHEP-89-31",
    doi = "10.1103/PhysRevLett.65.957",
    journal = "Phys. Rev. Lett.",
    volume = "65",
    pages = "957--959",
    year = "1990"
}

@article{DES:2020fxi,
    author = "Nadler, E. O. and others",
    collaboration = "DES",
    title = "{Milky Way Satellite Census. III. Constraints on Dark Matter Properties from Observations of Milky Way Satellite Galaxies}",
    eprint = "2008.00022",
    archivePrefix = "arXiv",
    primaryClass = "astro-ph.CO",
    reportNumber = "FERMILAB-PUB-20-277-AE, SLAC-PUB-17554, DES-2020-546",
    doi = "10.1103/PhysRevLett.126.091101",
    journal = "Phys. Rev. Lett.",
    volume = "126",
    pages = "091101",
    year = "2021"
}

@article{Nadler:2019zrb,
    author = "Nadler, Ethan O. and Gluscevic, Vera and Boddy, Kimberly K. and Wechsler, Risa H.",
    title = "{Constraints on Dark Matter Microphysics from the Milky Way Satellite Population}",
    eprint = "1904.10000",
    archivePrefix = "arXiv",
    primaryClass = "astro-ph.CO",
    doi = "10.3847/2041-8213/ab1eb2",
    journal = "Astrophys. J. Lett.",
    volume = "878",
    number = "2",
    pages = "32",
    year = "2019",
    note = "[Erratum: Astrophys.J.Lett. 897, L46 (2020), Erratum: Astrophys.J. 897, L46 (2020)]"
}

@article{mcclure-griffiths_atomic_2013,
    title = {{ATOMIC} {HYDROGEN} {IN} {A} {GALACTIC} {CENTER} {OUTFLOW}},
    volume = {770},
    copyright = {http://iopscience.iop.org/info/page/text-and-data-mining},
    issn = {2041-8205, 2041-8213},
    url = {https://iopscience.iop.org/article/10.1088/2041-8205/770/1/L4},
    doi = {10.1088/2041-8205/770/1/L4},
    language = {en},
    number = {1},
    urldate = {2026-08-18},
    journal = {The Astrophysical Journal},
    author = {McClure-Griffiths, N. M. and Green, J. A. and Hill, A. S. and Lockman, F. J. and Dickey, J. M. and Gaensler, B. M. and Green, A. J.},
    month = may,
    year = {2013},
    pages = {L4},
}

@ARTICLE{1992ApJ...396..649T,
       author = {{Timmes}, F.~X. and {Woosley}, S.~E.},
        title = "{The Conductive Propagation of Nuclear Flames. I. Degenerate C + O and O + NE + MG White Dwarfs}",
      journal = {\apj},
         year = 1992,
        month = sep,
       volume = {396},
        pages = {649},
          doi = {10.1086/171746},
       adsurl = {https://ui.adsabs.harvard.edu/abs/1992ApJ...396..649T}
}

@article{Sidhu_2020,
   title={Reconsidering astrophysical constraints on macroscopic dark matter},
   volume={101},
   ISSN={2470-0029},
   url={http://dx.doi.org/10.1103/PhysRevD.101.083503},
   DOI={10.1103/physrevd.101.083503},
   number={8},
   journal={Physical Review D},
   publisher={American Physical Society (APS)},
   author={Sidhu, Jagjit Singh and Starkman, Glenn D.},
   year={2020},
   month=Apr }

@article{carr_primordial_2017,
    title = {Primordial black hole constraints for extended mass functions},
    volume = {96},
    issn = {2470-0010, 2470-0029},
    url = {http://arxiv.org/abs/1705.05567},
    doi = {10.1103/PhysRevD.96.023514},
    language = {en},
    number = {2},
    urldate = {2026-08-05},
    journal = {Physical Review D},
    author = {Carr, Bernard and Raidal, Martti and Tenkanen, Tommi and Vaskonen, Ville and Veermäe, Hardi},
    month = jul,
    year = {2017},
    note = {arXiv:1705.05567 [astro-ph.CO]},
    pages = {023514},
}

@article{bellomo_primordial_2018,
    title = {Primordial {Black} {Holes} as {Dark} {Matter}: {Converting} {Constraints} from {Monochromatic} to {Extended} {Mass} {Distributions}},
    volume = {2018},
    issn = {1475-7516},
    shorttitle = {Primordial {Black} {Holes} as {Dark} {Matter}},
    url = {http://arxiv.org/abs/1709.07467},
    doi = {10.1088/1475-7516/2018/01/004},
    language = {en},
    number = {01},
    urldate = {2026-08-05},
    journal = {Journal of Cosmology and Astroparticle Physics},
    author = {Bellomo, Nicola and Bernal, José Luis and Raccanelli, Alvise and Verde, Licia},
    month = jan,
    year = {2018},
    note = {arXiv:1709.07467 [astro-ph.CO]},
    pages = {004--004},
}

@article{paczynski_gravitational_1986,
    title = {Gravitational microlensing by the galactic halo},
    volume = {304},
    issn = {0004-637X, 1538-4357},
    url = {http://adsabs.harvard.edu/doi/10.1086/164140},
    doi = {10.1086/164140},
    language = {en},
    urldate = {2026-08-19},
    journal = {The Astrophysical Journal},
    author = {Paczynski, B.},
    month = may,
    year = {1986},
    pages = {1},
}

@article{sidhu_macroscopic_2019,
    title = {Macroscopic dark matter constraints from bolide camera networks},
    volume = {100},
    issn = {2470-0010, 2470-0029},
    url = {https://link.aps.org/doi/10.1103/PhysRevD.100.123008},
    doi = {10.1103/PhysRevD.100.123008},
    language = {en},
    number = {12},
    urldate = {2026-08-19},
    journal = {Physical Review D},
    author = {Sidhu, Jagjit Singh and Starkman, Glenn},
    month = dec,
    year = {2019},
    pages = {123008},
}

@article{Graham_2015,
	doi = {10.1103/physrevd.92.063007},
  
	url = {https://doi.org/10.1103%2Fphysrevd.92.063007},
  
	year = 2015,
	month = {sep},
  
	publisher = {American Physical Society ({APS})},
  
	volume = {92},
  
	number = {6},
  
	author = {Peter W. Graham and Surjeet Rajendran and Jaime Varela},
  
	title = {Dark matter triggers of supernovae},
  
	journal = {Physical Review D}
}

@article{Acevedo_2019,
	doi = {10.1103/physrevd.100.043020},
  
	url = {https://doi.org/10.1103%2Fphysrevd.100.043020},
  
	year = 2019,
	month = {aug},
  
	publisher = {American Physical Society ({APS})},
  
	volume = {100},
  
	number = {4},
  
	author = {Javier F. Acevedo and Joseph Bramante},
  
	title = {Supernovae sparked by dark matter in white dwarfs},
  
	journal = {Physical Review D}
}

@article{Graham_2018,
	doi = {10.1103/physrevd.98.115027},
  
	url = {https://doi.org/10.1103%2Fphysrevd.98.115027},
  
	year = 2018,
	month = {dec},
  
	publisher = {American Physical Society ({APS})},
  
	volume = {98},
  
	number = {11},
  
	author = {Peter W. Graham and Ryan Janish and Vijay Narayan and Surjeet Rajendran and Paul Riggins},
  
	title = {White dwarfs as dark matter detectors},
  
	journal = {Physical Review D}
}

@article{Bramante_2015,
	doi = {10.1103/physrevlett.115.141301},
  
	url = {https://doi.org/10.1103%2Fphysrevlett.115.141301},
  
	year = 2015,
	month = {sep},
  
	publisher = {American Physical Society ({APS})},
  
	volume = {115},
  
	number = {14},
  
	author = {Joseph Bramante},
  
	title = {Dark Matter Ignition of Type Ia Supernovae},
  
	journal = {Physical Review Letters}
}

@article{Janish_2019,
	doi = {10.1103/physrevd.100.035008},
  
	url = {https://doi.org/10.1103%2Fphysrevd.100.035008},
  
	year = 2019,
	month = {aug},
  
	publisher = {American Physical Society ({APS})},
  
	volume = {100},
  
	number = {3},
  
	author = {Ryan Janish and Vijay Narayan and Paul Riggins},
  
	title = {Type Ia supernovae from dark matter core collapse},
  
	journal = {Physical Review D}
}

@article{Diamond_2022,
	doi = {10.1007/jhep03(2022)157},
  
	url = {https://doi.org/10.1007%2Fjhep03%282022%29157},
  
	year = 2022,
	month = {mar},
  
	publisher = {Springer Science and Business Media {LLC}
},
  
	volume = {2022},
  
	number = {3},
  
	author = {Melissa D. Diamond and David E. Kaplan},
  
	title = {Constraints on relic magnetic black holes},
  
	journal = {Journal of High Energy Physics}
}

@article{Acevedo_2022,
   title={Accelerating composite dark matter discovery with nuclear recoils and the Migdal effect},
   volume={105},
   ISSN={2470-0029},
   url={http://dx.doi.org/10.1103/PhysRevD.105.023012},
   DOI={10.1103/physrevd.105.023012},
   number={2},
   journal={Physical Review D},
   publisher={American Physical Society (APS)},
   author={Acevedo, Javier F. and Bramante, Joseph and Goodman, Alan},
   year={2022},
   month=jan }

@misc{raj2024supernovaesuperburstsdarkmatter,
      title={Supernovae and superbursts by dark matter clumps}, 
      author={Nirmal Raj},
      year={2024},
      eprint={2306.14981},
      archivePrefix={arXiv},
      primaryClass={hep-ph},
      url={https://arxiv.org/abs/2306.14981}, 
}

@article{Fedderke_2020,
	doi = {10.1103/physrevd.101.115021},
  
	url = {https://doi.org/10.1103%2Fphysrevd.101.115021},
  
	year = 2020,
	month = {jun},
  
	publisher = {American Physical Society ({APS})},
  
	volume = {101},
  
	number = {11},
  
	author = {Michael A. Fedderke and Peter W. Graham and Surjeet Rajendran},
  
	title = {White dwarf bounds on charged massive particles},
  
	journal = {Physical Review D}
}

@article{Acevedo_2021,
   title={Nuclear fusion inside dark matter},
   volume={103},
   ISSN={2470-0029},
   url={http://dx.doi.org/10.1103/PhysRevD.103.123022},
   DOI={10.1103/physrevd.103.123022},
   number={12},
   journal={Physical Review D},
   publisher={American Physical Society (APS)},
   author={Acevedo, Javier F. and Bramante, Joseph and Goodman, Alan},
   year={2021},
   month=jun }

@inbook{chandrasekhar,
  author    = "Chandrasekhar, Subrahmanyan",
  title     = " An introduction to the study of stellar structure. Vol. 2",
  chapter   = "Capter 11",
  publisher = "Courier Corporation",
  year      = "1958"
}

@article{derocco_dark_2025,
    title = {Dark wounds on icy moons: {Ganymede}'s subsurface ocean as a dark matter detector},
    volume = {112},
    issn = {2470-0010, 2470-0029},
    shorttitle = {Dark wounds on icy moons},
    url = {http://arxiv.org/abs/2508.00054},
    doi = {10.1103/d92d-9jwz},
    language = {en},
    number = {11},
    urldate = {2026-08-24},
    journal = {Physical Review D},
    author = {DeRocco, William},
    month = dec,
    year = {2025},
    note = {arXiv:2508.00054 [hep-ph]},
    pages = {115023},
}

@misc{uitenbroek_first_2026,
    title = {First {Search} for {Ultraheavy} {Dark} {Matter} {Using} a {Magnetically} {Levitated} {Particle}},
    url = {http://arxiv.org/abs/2608.20464},
    doi = {10.48550/arXiv.2608.20464},
    language = {en},
    urldate = {2026-08-24},
    publisher = {arXiv},
    author = {Uitenbroek, Dennis G. and Amaral, Dorian W. P. and Qin, Juehang and Langendorff, Jurriaan and Gingerich, Andrew and Oosterkamp, Tjerk H. and Tunnell, Christopher D.},
    month = aug,
    year = {2026},
    note = {arXiv:2608.20464 [hep-ph]},
}

@article{Caloni:2021bwp,
    author = "Caloni, Luca and Gerbino, Martina and Lattanzi, Massimiliano",
    title = "{Updated cosmological constraints on Macroscopic Dark Matter}",
    eprint = "2105.13932",
    archivePrefix = "arXiv",
    primaryClass = "astro-ph.CO",
    doi = "10.1088/1475-7516/2021/07/027",
    journal = "JCAP",
    volume = "07",
    pages = "027",
    year = "2021"
}

@article{SinghSidhu:2019nmh,
    author = "Singh Sidhu, Jagjit",
    title = "{Charge Constraints of Macroscopic Dark Matter}",
    eprint = "1912.04732",
    archivePrefix = "arXiv",
    primaryClass = "astro-ph.CO",
    doi = "10.1103/PhysRevD.101.043526",
    journal = "Phys. Rev. D",
    volume = "101",
    number = "4",
    pages = "043526",
    year = "2020"
}

@article{LHAASO:2024upb,
    author = "Cao, Zhen and others",
    collaboration = "LHAASO",
    title = "{Constraints on Ultraheavy Dark Matter Properties from Dwarf Spheroidal Galaxies with LHAASO Observations}",
    eprint = "2406.08698",
    archivePrefix = "arXiv",
    primaryClass = "astro-ph.HE",
    doi = "10.1103/PhysRevLett.133.061001",
    journal = "Phys. Rev. Lett.",
    volume = "133",
    number = "6",
    pages = "061001",
    year = "2024"
}

@article{CDEX:2025gtl,
    author = "Wang, Y. F. and others",
    collaboration = "CDEX",
    title = "{Constraints on ultraheavy dark matter from the CDEX-10 experiment at the China Jinping Underground Laboratory}",
    eprint = "2510.21458",
    archivePrefix = "arXiv",
    primaryClass = "hep-ex",
    doi = "10.1103/19fr-wcty",
    journal = "Phys. Rev. D",
    volume = "113",
    number = "5",
    pages = "052011",
    year = "2026"
}

@article{Liang:2026tjs,
    author = "Liang, Xunyu and Zhitnitsky, Ariel",
    title = "{Unidentified falling objects in the LHC as dark matter signals}",
    eprint = "2602.10562",
    archivePrefix = "arXiv",
    primaryClass = "hep-ph",
    doi = "10.1103/dm21-8tvz",
    journal = "Phys. Rev. D",
    volume = "114",
    number = "3",
    pages = "036001",
    year = "2026"
}

@article{Jiang:2025xln,
    author = "Jiang, Siyu and Yang, Aidi and Huang, Fa Peng",
    title = "{Macroscopic Dark Matter under siege: from White Dwarf Data to Gravitational Wave Detection}",
    eprint = "2511.23263",
    archivePrefix = "arXiv",
    primaryClass = "astro-ph.HE",
    month = "11",
    year = "2025"
}

@article{Bellinger:2025hrg,
    author = "Bellinger, Earl Patrick and Caplan, Matt E.",
    title = "{The Sun{\textquoteright}s Dark Core: Helioseismic and Neutrino Flux Constraints on a Compact Solar Center}",
    eprint = "2505.01503",
    archivePrefix = "arXiv",
    primaryClass = "astro-ph.SR",
    doi = "10.3847/1538-4357/ade70f",
    journal = "Astrophys. J.",
    volume = "988",
    number = "2",
    pages = "212",
    year = "2025"
}

@article{Miller:2025yyx,
    author = "Miller, Andrew L.",
    title = "{Gravitational wave probes of particle dark matter: A review}",
    eprint = "2503.02607",
    archivePrefix = "arXiv",
    primaryClass = "astro-ph.HE",
    doi = "10.1142/S0218271825300058",
    journal = "Int. J. Mod. Phys. D",
    volume = "35",
    number = "01",
    pages = "2530005",
    year = "2026"
}

@article{Kajino:2023wwa,
    author = "Kajino, Fumiyoshi",
    title = "{DIMS Experiment for Macroscopic Dark Matter and Interstellar Meteoroid Study}",
    doi = "10.22323/1.444.1376",
    journal = "PoS",
    volume = "ICRC2023",
    pages = "1376",
    year = "2023"
}

@article{Cooray:2021dvp,
    author = "Cooray, Vernon and Cooray, Gerald and Rubinstein, Marcos and Rachidi, Farhad",
    title = "{Could macroscopic dark matter (macros) give rise to mini-lightning flashes out of a blue sky without clouds?}",
    eprint = "2107.05338",
    archivePrefix = "arXiv",
    primaryClass = "physics.ao-ph",
    doi = "10.3390/atmos12091230",
    month = "7",
    year = "2021"
}

@article{Starkman:2022kft,
    author = "Starkman, Nathaniel and Starkman, Glenn D. and Winch, Harrison and Sidhu, Jagjit Singh",
    title = "{A Straight Lightning Bolt?! Observation of a Predicted Macro Dark Matter Signature}",
    eprint = "2202.12428",
    archivePrefix = "arXiv",
    primaryClass = "astro-ph.GA",
    month = "2",
    year = "2022"
}

@article{Bai:2022nsv,
    author = "Bai, Yang and Berger, Joshua and Korwar, Mrunal",
    title = "{IceCube at the frontier of macroscopic dark matter direct detection}",
    eprint = "2206.07928",
    archivePrefix = "arXiv",
    primaryClass = "hep-ph",
    doi = "10.1007/JHEP11(2022)079",
    journal = "JHEP",
    volume = "11",
    pages = "079",
    year = "2022"
}

@article{Ebadi:2021cte,
    author = "Ebadi, Reza and others",
    title = "{Ultraheavy dark matter search with electron microscopy of geological quartz}",
    eprint = "2105.03998",
    archivePrefix = "arXiv",
    primaryClass = "hep-ph",
    doi = "10.1103/PhysRevD.104.015041",
    journal = "Phys. Rev. D",
    volume = "104",
    number = "1",
    pages = "015041",
    year = "2021"
}

@article{Baum:2023cct,
    author = "Baum, Sebastian and others",
    title = "{Mineral detection of neutrinos and dark matter. A whitepaper}",
    eprint = "2301.07118",
    archivePrefix = "arXiv",
    primaryClass = "astro-ph.IM",
    reportNumber = "FERMILAB-PUB-23-501-SQMS-V",
    doi = "10.1016/j.dark.2023.101245",
    journal = "Phys. Dark Univ.",
    volume = "41",
    pages = "101245",
    year = "2023"
}

@article{Bramante:2018tos,
    author = "Bramante, Joseph and Broerman, Benjamin and Kumar, Jason and Lang, Rafael F. and Pospelov, Maxim and Raj, Nirmal",
    title = "{Foraging for dark matter in large volume liquid scintillator neutrino detectors with multiscatter events}",
    eprint = "1812.09325",
    archivePrefix = "arXiv",
    primaryClass = "hep-ph",
    doi = "10.1103/PhysRevD.99.083010",
    journal = "Phys. Rev. D",
    volume = "99",
    number = "8",
    pages = "083010",
    year = "2019"
}

@article{Diamond:2021dth,
    author = "Diamond, Melissa D. and Kaplan, David E. and Rajendran, Surjeet",
    title = "{Binary collisions of dark matter blobs}",
    eprint = "2112.09147",
    archivePrefix = "arXiv",
    primaryClass = "hep-ph",
    reportNumber = "FERMILAB-PUB-21-0883-SQMS-V",
    doi = "10.1007/JHEP01(2023)136",
    journal = "JHEP",
    volume = "01",
    pages = "136",
    year = "2023"
}

@article{SinghSidhu:2018oqs,
    author = "Singh Sidhu, Jagjit and Abraham, Roshan Mammen and Covault, Corbin and Starkman, Glenn",
    title = "{Macro detection using fluorescence detectors}",
    eprint = "1808.06978",
    archivePrefix = "arXiv",
    primaryClass = "astro-ph.HE",
    doi = "10.1088/1475-7516/2019/02/037",
    journal = "JCAP",
    volume = "02",
    pages = "037",
    year = "2019"
}

@article{Herrin:2005kb,
    author = "Herrin, Eugene T. and Rosenbaum, Doris C. and Teplitz, Vigdor L.",
    title = "{Seismic search for strange quark nuggets}",
    eprint = "astro-ph/0505584",
    archivePrefix = "arXiv",
    doi = "10.1103/PhysRevD.73.043511",
    journal = "Phys. Rev. D",
    volume = "73",
    pages = "043511",
    year = "2006"
}

@article{Cyncynates:2016rij,
    author = "Cyncynates, David and Chiel, Joshua and Sidhu, Jagjit and Starkman, Glenn D.",
    title = "{Reconsidering seismological constraints on the available parameter space of macroscopic dark matter}",
    eprint = "1610.09680",
    archivePrefix = "arXiv",
    primaryClass = "astro-ph.CO",
    doi = "10.1103/PhysRevD.95.063006",
    journal = "Phys. Rev. D",
    volume = "95",
    number = "6",
    pages = "063006",
    year = "2017",
    note = "[Addendum: Phys.Rev.D 95, 129903 (2017)]"
}

@article{Bramante:2021dyx,
    author = "Bramante, Joseph and Kavanagh, Bradley J. and Raj, Nirmal",
    title = "{Scattering Searches for Dark Matter in Subhalos: Neutron Stars, Cosmic Rays, and Old Rocks}",
    eprint = "2109.04582",
    archivePrefix = "arXiv",
    primaryClass = "hep-ph",
    doi = "10.1103/PhysRevLett.128.231801",
    journal = "Phys. Rev. Lett.",
    volume = "128",
    number = "23",
    pages = "231801",
    year = "2022"
}

@article{Dessert:2021wjx,
    author = "Dessert, Christopher and Johnson, Zachary",
    title = "{Red-giant branch stellar cores as macroscopic dark matter detectors}",
    eprint = "2112.06949",
    archivePrefix = "arXiv",
    primaryClass = "hep-ph",
    doi = "10.1103/PhysRevD.106.103034",
    journal = "Phys. Rev. D",
    volume = "106",
    number = "10",
    pages = "103034",
    year = "2022"
}

@article{Das:2021drz,
    author = "Das, Anirban and Ellis, Sebastian A. R. and Schuster, Philip C. and Zhou, Kevin",
    title = "{Stellar Shocks from Dark Matter Asteroid Impacts}",
    eprint = "2106.09033",
    archivePrefix = "arXiv",
    primaryClass = "hep-ph",
    reportNumber = "SLAC-PUB-17604",
    doi = "10.1103/PhysRevLett.128.021101",
    journal = "Phys. Rev. Lett.",
    volume = "128",
    number = "2",
    pages = "021101",
    year = "2022"
}

@article{Shvartzvald:2023ofi,
    author = "Shvartzvald, Y. and others",
    title = "{ULTRASAT: A Wide-field Time-domain UV Space Telescope}",
    eprint = "2304.14482",
    archivePrefix = "arXiv",
    primaryClass = "astro-ph.IM",
    doi = "10.3847/1538-4357/ad2704",
    journal = "Astrophys. J.",
    volume = "964",
    number = "1",
    pages = "74",
    year = "2024"
}

@article{Cleaver:2025etu,
    author = "Cleaver, Damon and McCabe, Christopher and O'Hare, Ciaran A. J.",
    title = "{Listening for ultraheavy dark matter with underwater acoustic detectors}",
    eprint = "2502.17593",
    archivePrefix = "arXiv",
    primaryClass = "hep-ph",
    reportNumber = "KCL-PH-TH/2025-04",
    doi = "10.1103/jpzr-msx1",
    journal = "Phys. Rev. D",
    volume = "112",
    number = "6",
    pages = "063060",
    year = "2025"
}

\end{document}